\documentclass[showpacs,preprintnumbers,eqsecnum,twocolumn,prd,floatfix,nofootinbib]{revtex4-1}  

\usepackage{graphicx}
\usepackage{latexsym}
\usepackage{amsmath}
\usepackage{amssymb}

\def\beq{\begin{equation}}
\def\eeq{\end{equation}}

\begin{document}

\title{Spin-dependent two-body interactions from gravitational self-force computations}

\author{Donato \surname{Bini}$^1$}
\author{Thibault \surname{Damour}$^2$}
\author{Andrea \surname{Geralico}$^1$}

\affiliation{$^1$Istituto per le Applicazioni del Calcolo ``M. Picone'', CNR, I-00185 Rome, Italy\\
$^2$Institut des Hautes Etudes Scientifiques, 91440 Bures-sur-Yvette, France}

\date{\today}

\begin{abstract}
We analytically compute, through the eight-and-a-half post-Newtonian order and  the fourth-order in spin, the gravitational self-force correction to Detweiler's gauge invariant redshift function for a small mass in circular orbit around a Kerr black hole. Using the first law of mechanics for black hole binaries with spin [L.~Blanchet, A.~Buonanno and A.~Le Tiec,
  Phys.\ Rev.\ D {\bf 87}, 024030 (2013)] we transcribe our results into a knowledge of various spin-dependent couplings, as encoded within the spinning effective-one-body model of T.~Damour and A.~Nagar  [Phys.\ Rev.\ D {\bf 90}, 044018 (2014)].
We also compare our analytical results to the (corrected) numerical self-force results of A.~G.~Shah, J.~L.~Friedman and T.~S.~Keidl [Phys.\ Rev.\ D {\bf 86}, 084059 (2012)], from which we show how to directly extract physically relevant spin-dependent couplings. 
\end{abstract}

\maketitle

\section{Introduction}
The imminent prospect of detecting gravitational-wave signals emitted by inspiralling and coalescing binary systems gives a strong motivation for improving our {\it analytical} knowledge of the general relativistic dynamics of two-body  systems. 
The effective one-body (EOB) formalism \cite{Buonanno:1998gg,Buonanno:2000ef,Damour:2000we,Damour:2001tu} has established itself as   the most accurate way  of theoretically describing the dynamics of inspiralling and coalescing compact binary systems. 

Recent years have witnessed a useful synergy between EOB theory and various other analytical-relativity approaches to the two-body problem, notably post-Newtonian (PN) theory \cite{Schafer:2009dq,Blanchet:2013haa} and gravitational self-force (GSF) theory \cite{Barack:2009ux,Poisson:2011nh}. Several different flavors of GSF theory have been useful in this respect: numerical GSF computations, analytical GSF computations and, more recently, mixed numerical-analytical GSF computations \cite{Detweiler:2008ft,Blanchet:2009sd,Damour:2009sm,Blanchet:2010zd,Barack:2010ny,LeTiec:2011dp,Barausse:2011dq,Akcay:2012ea,Shah:2013uya,Bini:2014nfa,Dolan:2013roa,Keidl:2010pm,Shah:2012gu,Bini:2013zaa,Bini:2013rfa,Bini:2014ica,Dolan:2014pja,Bini:2014zxa,Bini:2015bla,Bini:2015mza,Kavanagh:2015lva}
In addition, numerical relativity simulations have also been of crucial importance for completing the current analytical knowledge in EOB theory \cite{Buonanno:2009qa,LeTiec:2011bk,Damour:2011fu,Damour:2012ky,Taracchini:2013rva,Hinder:2013oqa,Pan:2013rra,Pan:2013tva,Damour:2014afa,Damour:2014sva}.

The aim of the present work is to extract new information about {\it spin-dependent} two-body~\footnote{We denote the masses and the spins of the two-body system as $m_1$, $m_2$, $S_1=m_1 a_1=m_1^2\hat a_1$ and $S_2=m_2 a_2=m_2^2\hat a_2$ (using units where $G=c=1$). We follow here the convention $m_1\le m_2$ so that $X_1=\frac{m_1}{m_1+m_2}=\frac12 (1-\sqrt{1-4\nu})$, $X_2=\frac{m_2}{m_1+m_2}=\frac12 (1+\sqrt{1-4\nu})$ where $\nu=m_1m_2/(m_1+m_2)^2$.} interactions (as encoded in the EOB Hamiltonian) from both analytical and numerical GSF computations of Detweiler's perturbed redshift function
$\delta U(m_2\Omega, \hat a_2)$ \cite{Detweiler:2008ft}  around a spinning (Kerr) black hole of mass $m_2$ and Kerr parameter $a_2=m_2 \, \hat a_2$. 

A formalism for computing the $O(m_1)$ GSF correction to the redshift, $\delta U(m_2\Omega, \hat a_2)$ (where $\Omega$ denotes the orbital frequency of the small mass $m_1$ in circular orbit around a Kerr black hole of mass $m_2$ and spin $S_2=m_2^{\, 2} \,  \hat a_2$) has  been set up in Refs.~\cite{Keidl:2010pm,Shah:2012gu}.
Our high PN order analytical calculations of $\delta U$, whose results we shall present below, have been based on the formalism of \cite{Keidl:2010pm,Shah:2012gu} together with an extension of the techniques we have recently developed \cite{Bini:2013zaa,Bini:2013rfa} for efficiently computing the PN expansion of various gauge invariant GSF functions in the case where the small mass $m_1$  orbits a non-spinning black hole. While we were deriving our results we were informed by A. Shah \cite{priv_com} of the existence of parallel work by him and his collaborators leading also to high PN order computations of $\delta U$. Some of their results have been recently presented in various conferences \cite{shah_capra2015,shah_MG14} while we were finalizing our calculations. There is a complete agreement between their current results and the more  accurate (in PN order) and more complete (in order of expansion in $\hat a_2$) ones that we present below. Moreover, Ref. \cite{Shah:2012gu} has published numerical GSF data on the function
$\delta U(m_2\Omega, \hat a_2)$ especially in the strong field domain. [Actually, the published data were marred by a  minor technical error, but A. Shah kindly provided us with a corrected version of the data  of \cite{Shah:2012gu}; see also \cite{vandeMeent:2015lxa}]. We will show below how these numerical data can be used to complement the weak-field knowledge given by the PN expansion of $\delta U(m_2\Omega, \hat a_2)$ (to be discussed next) by giving access to some
strong-field information.

A central tool allowing one to relate Detweiler's redshift function to the Hamiltonian of a two-body system is the so-called \lq\lq first law of binary black hole mechanics" \cite{LeTiec:2011ab,Blanchet:2012at,Tiec:2015cxa}. We shall show below how to use the first law of spinning binaries \cite{Blanchet:2012at} to transcribe the information contained in the function $\delta U (m_2 \Omega , \hat a_2)$ into a knowledge of the spin-dependent couplings as encoded within the spinning EOB formalism of Ref.~\cite{Damour:2014sva}. More precisely, we shall show (generalizing Ref.~\cite{Barausse:2011dq}) how to algebraically extract from $\delta U$ the first order GSF corrections to two EOB potentials: (i) the radial equatorial potential $A(r , m_1 , m_2 , S_1 , S_2)$; and (ii) the main (``$S$-type'') spin-orbit coupling potential $G_S (r , m_1 , m_2 , S_1 , S_2)$. [The first-order GSF correction to the second (``$S_*$-type'') spin-orbit coupling $G_{S_*} (r , m_1 , m_2 , S_1 , S_2)$ has been recently extracted from other GSF computations in Refs.~\cite{Bini:2014ica}.]

\section{Connection between the GSF correction to the redshift and the $O(m_1^{\, 2})$ correction to the two-body Hamiltonian}

The first law of binary black hole mechanics \cite{LeTiec:2011ab,Blanchet:2012at,Tiec:2015cxa} allows one to relate Detweiler's redshift function to the Hamiltonian of a two-body system. In the circular case (to which we shall restrict ourselves here), the latter first law equates (to linear order in $S_1$) the redshift of particle 1,
\beq
\label{z1}
z_1=\frac{1}{U}=\left( \frac{ds}{dt} \right)_1
\eeq
to the partial derivative of the total (two-body) Hamiltonian (including the contribution of the rest masses, $m_1+m_2$)  with respect to (wrt)  $m_1$:
\beq
\label{1stlaw}
\frac{1}{U}=z_1=\frac{\partial H}{\partial m_1}(r,P_\phi,m_1,m_2,S_1,S_2) + O(S_1^2) \,.
\eeq
Note that the dynamical variables~\footnote{For the considered circular, equatorial, parallel-spin case, where $P_r=0$, $\theta=\frac{\pi}{2}$, $P_\theta=0$.} $r,P_\phi,S_1,S_2$ are kept fixed when differentiating wrt $m_1$. [Here $P_i=(P_r,P_\theta,P_\phi)$ denote the canonical momenta of the relative position vector in the center of mass frame, so that $P_\phi$ is the total orbital angular momentum of the system.]

In the following, we shall restrict ourselves to the case $S_1=0$. Eq. (\ref{1stlaw}) then yields $z_1$ (or, equivalently, $U=1/z_1$) as a function of dynamical variables: $z_1(r,P_\phi)$. By contrast, GSF computations give access to the self-force correction  $\delta z_1$ to $z_1$ (or, equivalently, $\delta U=-\delta z_1/z_1^2$), considered as a function of the dimensionless frequency parameter
\beq
\label{y}
y=(m_2\Omega)^{2/3}\,
\eeq
(and of the dimensionless spin parameter $\hat a_2=S_2/m_2^2$). As $\Omega$ is the partial derivative of $H$ wrt  orbital angular momentum
\beq
\label{Omega}
\Omega= \frac{\partial H}{\partial P_\phi}(r,P_\phi,m_1,m_2,S_2)\,,
\eeq
the passage from the variable $P_\phi$ to the variable $\Omega$ 
or, more completely,  from the pair of variables $(r,P_\phi)$ to the pair of variables $(e_r,\Omega)$, where
\beq
\label{er}
e_r\equiv \frac{\partial H}{\partial r}(r,P_\phi,m_1,m_2,S_2)
\eeq
denotes the radial equation of motion (namely, $e_r=0$ along circular orbits), is conveniently associated with the following Legendre transform of the Hamiltonian (where the ellipsis denote $m_2$ and $S_2$, or, equivalently, $m_2$ and $\hat{a}_2$)
\begin{widetext}
\beq
\widetilde H(e_r, \Omega, m_1,\ldots):=\left[H(r,P_\phi,m_1,\ldots)-P_\phi  \frac{\partial H}{\partial P_\phi}(r,P_\phi,m_1,\ldots)-r \frac{\partial H}{\partial r}(r,P_\phi,m_1, \ldots) \right]_{\stackrel{P_\phi=P_\phi(\Omega, e_r)}{r=r(\Omega, e_r)}} \,.
\eeq
\end{widetext}
Let us apply this Legendre transform to a two-body Hamiltonian of the form
\begin{eqnarray}
H(r,P_\phi,m_1,\ldots)&=&m_2 +m_1 H^{(m_1)}(r,p_\phi,\ldots)\\
&+&m_1^2 H^{(m_1^2)}(r,p_\phi,\ldots)+O(m_1^3)\,,\nonumber
\end{eqnarray}
where we introduced the following rescaled (and dimensionless) angular momentum
\beq
p_\phi=\frac{P_\phi}{m_1 m_2}\,.
\eeq
It is then easily found that, along circular orbits (i.e., for $e_r=0$), and for a given value of the dimensionless frequency parameter \eqref{y}, we have the simple link
\beq
\label{link}
\frac12 \delta z_1 (y, \ldots) = m_1 H^{(m_1^2)}(r,p_\phi,\ldots) \bigg|_{
\stackrel{\frac{\partial}{\partial r} H^{(m_1)}(r,p_\phi,\ldots)=0,} 
{\frac{\partial }{\partial p_\phi}H^{(m_1)}(r,p_\phi,\ldots)=m_2\Omega
}}\,.
\eeq
In other words, the first-order GSF correction~\footnote{For the redshift, $z_1$ or its inverse $U=1/z_1$, considered as functions of $m_2\Omega$, we denote the $O(m_1)$ GSF correction (including its factor $m_1$) by a $\delta$, so that, e.g.,
$z_1(m_2\Omega,m_1)=z_1(m_2\Omega,0)+\delta z_1(m_2\Omega)+O(m_1^2)$.} (as function of $m_2\Omega$) 
\begin{eqnarray}
\label{del_z1}
\delta z_1(y,m_2,S_2)&\equiv& -z_1^2 \delta U(y,m_2,S_2)\nonumber\\
&=&-\frac{\delta U(y,m_2,S_2)}{U^2}
\end{eqnarray}
is simply numerically equal (along circular orbits) to (twice) the value of the $O(m_1^2)$ contribution  to the two-body Hamiltonian (as a function of the  variables  $r$, $p_\phi$, $m_2$ and $S_2$). This simple algebraic link (which does not involve any differentiation of $H^{(m_1^2)}$) generalizes to the spinning case the algebraic link between $\delta z_1$ and the $O(\nu)$ contribution to the EOB radial potential found in Ref. \cite{Barausse:2011dq}.

In view of the usefulness of the EOB formalism for describing the interaction of two-body systems, we shall transcribe the simple link (\ref{link}) in terms of the building blocks of the EOB Hamiltonian.
We recall that the EOB Hamiltonian is first written as (with $M=m_1+m_2$, $\mu=m_1m_2/M$, $\nu=\mu/M$)
\beq
\label{EOB_ham}
H(r,P_r,P_\phi,m_1,m_2,S_1,S_2)=M\sqrt{1+2\nu \left(\frac{ H_{\rm eff}}{\mu}-1\right)}\,,
\eeq
where the effective Hamiltonian $H_{\rm eff}$ has the general form (here considered for the equatorial circular dynamics of parallel-spin binary systems) \cite{Damour:2007nc,Damour:2014sva}
\begin{eqnarray}
\label{Heff}
H_{\rm eff}(r,L,m_1,m_2,S_1,S_2)&=& G_S^{\rm phys} LS + G_{S_*}^{\rm phys} L S_* \nonumber\\
&+& \sqrt{A\left(\mu^2+\frac{L^2}{r_c^2}  \right)}\,.
\end{eqnarray}
Here $L=P_\phi$ is the total orbital angular momentum of the system, $S$ and $S_*$ are the following symmetric combinations of the two spins
\begin{eqnarray}
S&\equiv&S_1+S_2=m_1a_1+m_2a_2\nonumber\\
&=& m_1^2 \hat a_1 +m_2^2\hat a_2\,,\nonumber\\
S_*&\equiv& \frac{m_2}{m_1}S_1+\frac{m_1}{m_2}S_2=m_2a_1+m_1a_2\nonumber\\
&=& m_1m_2 (\hat a_1+\hat a_2)\,,
\end{eqnarray}
$r_c^2$ is the following (squared) \lq\lq centrifugal radius"
\beq
\label{rc_def}
r_c^2\equiv r^2 +a_0^2 \left(1+\frac{2M}{r}\right)
\eeq
with
\beq
a_0=a_1+a_2=m_1 \hat a_1 +m_2 \hat a_2\,,
\eeq
and where $G_{S}^{\rm phys}(r,m_1,m_2,S_1,S_2)$, $G_{S_*}^{\rm phys}(r,m_1,m_2,S_1,S_2)$  are two spin-orbit coupling functions.

The structure of the effective Hamiltonian, Eq. \eqref{Heff}, shows that the energetics of circular orbits can be encoded in three separate functions: two spin-orbit coupling functions, $G_S^{\rm phys}(r,m_1,m_2,S_1,S_2)$, $G_{S_*}^{\rm phys}(r,m_1,m_2,S_1,S_2)$ and one radial potential $A(r,m_1,m_2,S_1,S_2)$. All these functions a priori depend on the two spins $S_1$ and $S_2$. In addition, the spins enter the centrifugal term $L^2/r_c^2$
via the definition \eqref{rc_def} of $r_c^2$. [Following \cite{Balmelli:2015zsa} we have taken as effective Kerr parameter in $r_c^2$ the combination $a_0=a_1+a_2$ which encodes the leading-order (LO) spin-spin coupling \cite{Damour:2001tu}.] While the EOB radial potential is dimensionless, the spin-orbit coupling functions $G_S^{\rm phys}$, $G_{S_*}^{\rm phys}$ have dimension [length]$^{-3}$ or [mass]$^{-3}$ (in the units $G=c=1$ that we use).
It will be convenient in the following to work with the following dimensionless versions of these functions
\begin{eqnarray}
G_S(r,m_1,m_2,S_1,S_2)&\equiv& M^3 G_S^{\rm phys},\nonumber\\
G_{S_*}(r,m_1,m_2,S_1,S_2)&\equiv& M^3 G_{S_*}^{\rm phys}\,,
\end{eqnarray}
where $M=m_1+m_2$.

In the following, we shall restrict ourselves to the case where the spin $S_1$ vanishes. In that case, the dimensionless, $\mu$-rescaled effective Hamiltonian $\hat H_{\rm eff}\equiv H_{\rm eff}/\mu$ reads
\begin{eqnarray}
\label{H_eff_2}
\hat H_{\rm eff}(r,p_\phi,m_1,m_2,S_2)&=& G_S p_\phi X_2 \hat a_2+G_{S_*}p_\phi \nu \hat a_2\nonumber\\
&+& \sqrt{A(1+p_\phi^2 u_c^2)}\,,
\end{eqnarray}
where
\beq
u_c^2\equiv \left( \frac{M}{r_c} \right)^2\equiv \frac{u^2}{1+\hat a_0^2 u^2 (1+2u)}\,,
\eeq
with
\beq
u\equiv \frac{M}{r}\,, \quad \hat a_0\equiv \frac{a_0}{M}=X_2 \hat a_2\,.
\eeq
In view of the form  \eqref{EOB_ham} of the EOB Hamiltonian, the link, Eq. \eqref{link}, between the first-order GSF contribution $\delta z_1$ to the 
redshift and the $m_1$-expansion of the Hamiltonian, shows that $\delta z_1$ only depends on the first-order GSF contribution to $\hat H_{\rm eff}$, and therefore on the first-order contributions to $A$ and $G_S$. We parametrize the latter by decomposing $A$ and $G_S$ as
\begin{eqnarray}
A(r,m_1,m_2,S_2)&=& A^{\rm Kerr}(r,M,a_0)+\delta A +O(\nu^2)\,,\nonumber\\
G_S(r,m_1,m_2,S_2)&=& G_S^{\rm Kerr}(r,M,a_0)+\delta G_S +O(\nu^2)\,,\nonumber
\end{eqnarray}
where the consideration of the Hamiltonian of a non-spinning test-particle in a Kerr background (here conventionally taken of mass $M=m_1+m_2$, and Kerr parameter $a_0=a_1+a_2=a_2$) determines the zeroth-order GSF potentials as (see Appendix)
\begin{eqnarray}
\label{eq1.20}
A^{\rm Kerr}(r,M,a_0)&=& 1-2u+4\hat a_0^2 u^2 u_c^2\,, \nonumber\\
G_S^{\rm Kerr}(r,M,a_0)&=&2uu_c^2\,.
\end{eqnarray}

On the other hand, the presence of a factor $\nu$ in the $G_{S_*}$ contribution to $\hat H_{\rm eff}$, Eq. \eqref{H_eff_2}, implies that $\delta z_1$ only depends on the zeroth-order GSF contribution to $G_{S_*}$. The latter is determined (as emphasized in \cite{Barausse:2009xi}) by the spin-orbit coupling of a {\it spinning} test-particle in a Kerr background. Following Refs. \cite{Damour:2007nc}, the latter is most simply determined by computing the geodetic spin-precession rate. When considering, as we do here, equatorial circular orbits the spin-precession only depends on the equatorial restriction $(\theta =\pi/2)$ of the metric, say $ds_{\rm eq}^2= -A_{\rm eq}dt^2 + r_c^2 (d\phi -\omega_{\rm eq} dt)^2+B_{\rm eq}dr^2$.  Separating the spin-orbit contribution linked to the metric functions $A_{\rm eq} , r_c^2$ and $B_{\rm eq}$ from the spin-spin interaction linked to $\omega_{\rm eq}$ (i.e., calculating simply the spin precession with $\omega_{\rm eq} =0$), yields
\begin{eqnarray}
\label{GSs}
G_{S_*}^{\rm test-particle}&=&\left(\frac{M}{r_c}\right)^2 \left[\frac{r_c \nabla \sqrt{A_{\rm eq}}}{1+\sqrt{Q}} \right. \nonumber\\
&& \left. +\frac{(1-\nabla r_c)\sqrt{A_{\rm eq}} }{\sqrt{Q}} \right]+O(\nu)\,,
\end{eqnarray}
where $\nabla \equiv (B_{\rm eq})^{-1/2}d/dr$ is a proper radial gradient and $Q\equiv 1+P_\phi^2/(\mu r_c)^2$. [Here, we are in the small mass-ratio limit and we denoted, for simplicity, the large mass by $M$ and the small one by $\mu$. One could equivalently replace $M$ by $m_2$ and $\mu$ by $m_1$.] The explicit, relevant expression of $G_{S_*}^{\rm test-particle}$ in a Kerr background is given in the Appendix.

\smallskip

Using the above results, and notations, we derive the following explicit algebraic link between the GSF correction $\delta z_1$ to the redshift and the GSF corrections $\delta A$, $\delta G_S$ to the two relevant EOB potentials:
\begin{eqnarray}
\label{12deltaz1}
\frac12 \delta z_1(y,\hat a)& =& \left[ \frac{\delta A(u,\hat a)}{2z_1}+p_\phi \hat a \delta G_S(u,\hat a) \right. \nonumber\\
&&\left. \quad + \nu {\mathcal K}(u,\hat a) \right]_{u=y'(y)}\,.
\end{eqnarray}
In this result, which is valid only to order $O(\nu)$, $z_1$ and $p_\phi$ can be replaced by their (circular) test-mass limits (computed in a Kerr background).
The explicit expressions of $z_1^{\rm Kerr}$ and $p_\phi^{\rm Kerr}$ (as functions of $u$ and $\hat a$) are given  by
(see Appendix A)
\begin{eqnarray}
\label{Kerrpphi}
p_\phi^{\rm Kerr}&=&\frac{1-2\hat a_2 u^{3/2} +\hat a_2^2 u^2}{\sqrt{u} \sqrt{1-3u+2\hat a_2 u^{3/2}}} \,,\nonumber\\
z_1^{\rm Kerr}&=& \frac{\sqrt{1-3u+2\hat a_2 u^{3/2}}}{1+\hat a_2 u^{3/2}}\,.
\end{eqnarray}

To simplify the notation, we have denoted the dimensionless spin parameter entering the $O(\nu)$ corrections of Eq. \eqref{12deltaz1}, as $\hat a$ [It is equal to $\hat a_2=\lim_{\nu \to 0} \hat a_0$, the dimensionless Kerr parameter of the large mass.] In addition, the extra contribution $\nu {\mathcal K}(u,\hat a)$ in Eq. \eqref{12deltaz1} gathers the analytically known contributions to $\frac12 \delta z_1$ coming from various $O(m_1^{\,2})$ terms entering Eq. \eqref{EOB_ham} [coming from various occurrences of $M=m_1+m_2$ or $X_2=1-m_1/(m_1+m_2)$, and from the square-root nature of $H$ as a function of $1+2\nu (\hat H_{\rm eff}-1)$]. The latter known contribution is explicitly given by
\begin{eqnarray}
{\mathcal K}(u,\hat a)&=&\frac12 u  \frac{1-4u  \left(1-\frac12 \hat a \sqrt{u }  \right)^2}{1-3u +2\hat a  u ^{3/2}}+K(u , \hat a )\,,
\end{eqnarray}
where
\beq
K(u, \hat a)=\alpha p_\phi^2 +\beta p_\phi +\gamma\,,
\eeq
and
\begin{eqnarray}
\alpha
&=& z_1 u_{c}^2  (1-\phi)\,,\nonumber\\
\beta &=& \hat a  \left[  G_{S_*} + 2u u_{c }^2(1-2\phi) \right]\,,\nonumber\\
\gamma &=& -\frac{u }{z_1}(1-2\phi)^2\,,
\end{eqnarray}
with 
\beq
\phi=\hat a^2 u  u_{c}^2\,. 
\eeq
 
The (dimensionless) first-order GSF contributions $\delta A$ and $\delta G_S$ are, modulo a prefactor $\nu$, functions of $m_2/r$ and $\hat a_2$. As $\delta A$ and $\delta G_S$ are $O(\nu)$ corrections, one can denote their arguments (as we did for $\nu {\mathcal K}$) simply as $u$ and $\hat a$.

The last step which is needed for computing the function $\delta z_1 (y,\hat a)$ in terms of the functions $\delta A(u, \hat a)$ and $\delta G_S(u, \hat a)$ is to express the dimensionless gravitational potential $ u=m_2/r+O(\nu)$ as a function of the dimensionless parameter $y$, Eq. \eqref{y}. At the zeroth-order in $\nu$ where this transformation is needed, this follows from the known Kepler law around a Kerr black -hole (of mass $m_2$ and spin $S_2$), which is (see also the Appendix) 
\beq
\label{ucirc}
u^{\rm circ}=y'(y, \hat a)
\eeq
where the function $y'(y, \hat a)$ is defined as 
\beq
\label{eq1.27}
y'(y, \hat a)\equiv \frac{y}{(1-\hat a y^{3/2})^{2/3}} \,.
\eeq

Eq. \eqref{12deltaz1} is one of the main tools of the present paper. We will show in the following how to use it to extract both
$\delta A(u,\hat a)$ and $\delta G_S(u,\hat a)$ from the GSF calculation of $\delta z_1(y,\hat a)$, thereby furthering our current knowledge of spin-dependent interactions in binary systems.

\section{Analytical computation of the self-force correction to the redshift function around a Kerr black hole}

Detweiler \cite{Detweiler:2008ft} has pointed out the potential importance of computing the (gauge-invariant) first-order GSF correction $\delta U(\Omega)$ to the redshift function $\left(\frac{dt}{ds} \right)_1=U(\Omega)$, associated with the sequence of circular orbits of an extreme mass-ratio binary system $m_1\ll m_2$.
He pioneered the computation (both numerical and analytical) of $\delta U(\Omega)$ in the case where the large-mass body is a Schwarzschild black hole. Many works have extended his results to higher accuracy, and have generalized the redshift function to other gauge-invariant functions \cite{Blanchet:2009sd,Damour:2009sm,Blanchet:2010zd,Barack:2010ny,LeTiec:2011dp,Barausse:2011dq,Akcay:2012ea,Shah:2013uya,Bini:2014nfa,Dolan:2013roa,Keidl:2010pm,Shah:2012gu,Bini:2013zaa,Bini:2013rfa,Bini:2014ica,Dolan:2014pja,Bini:2014zxa,Bini:2015bla,Bini:2015mza,Kavanagh:2015lva}.

The generalization of the redshift function to the case where the large mass is a Kerr black hole poses significant technical challenges, which have been tackled in Refs. \cite{Keidl:2010pm,Shah:2012gu} by using a radiation gauge together with a Hertz potential approach. Here we apply to the approach of Refs. \cite{Keidl:2010pm,Shah:2012gu} the analytical techniques we have recently developed to compute the PN expansion of $\delta U$ in the Schwarzschild case \cite{Bini:2013zaa,Bini:2013rfa}. The generalization of our analytical approach to the Kerr case is conceptually straightforward (in view of the work of Refs. \cite{Mano:1996mf,Mano:1996vt}) but has necessitated quite a few new technical developments. We shall leave to future work a detailed explanation of the latter technical tools, and recall here only the basic conceptual aspects of our 
analytical approach, before giving our final results.

Computing the first-order GSF correction $\delta U(m_2 \Omega, \hat a_2)$ to the redshift function $\left(\frac{dt}{ds} \right)_1=U(m_2\Omega, \hat a_2)$ (or, equivalently, the correction $\delta z_1(m_2 \Omega, \hat a_2)$ to $z_1=\left(\frac{ds}{dt} \right)_1=1/U$) is equivalent  
to computing the 
{\it regularized} value, along the world line $y_1^\mu$ of the small mass $m_1$, of the double contraction of the $O(m_1)$ metric perturbation $h_{\mu\nu}$
\beq
g_{\mu\nu}(x; m_1, m_2, \hat a_2)=g_{\mu\nu}^{(0)}(x; m_2, \hat a_2)+h_{\mu\nu}(x)+O(m_1^2)\,,
\eeq
[where $g_{\mu\nu}^{(0)}(x; m_2, \hat a_2)$ is a Kerr metric of mass $m_2$ and spin $S_2\equiv m_2^2 \hat a_2$] with the helical Killing vector 
$k^\mu \partial_\mu=\partial_t +\Omega \partial_\phi$ [such that $u_1^\mu\equiv \frac{dy_1^\mu}{ds}=Uk^\mu$]. More precisely, the function
\beq
h_{kk}(m_2\Omega, \hat a_2)\equiv {\rm Reg}_{x\to y_1}[h_{\mu\nu}(x)k^\mu k^\nu]
\eeq
(computed in the mostly-plus signature) determines both
\beq
\delta U(m_2 \Omega, \hat a_2)=+ \frac12 \frac{h_{kk}}{z_1^3}
\eeq
and
\beq
\delta z_1(m_2 \Omega, \hat a_2)=-\frac12  \frac{h_{kk}}{z_1}\,.
\eeq
On the right-hand side (rhs) of these expressions $z_1$ denotes the zeroth-order redshift (computed for a test particle in Kerr), as given by the second Eq.~(\ref{Kerrpphi}).

\smallskip

In addition, the dimensionless gravitational potential $u=m_2/r_0+O(\nu)$ (where $r_0$ is the orbital radius) entering the latter expression must be expressed as a function of the dimensionless
orbital frequency $m_2\Omega$ by means of Eqs.~(\ref{ucirc}), \eqref{eq1.27}. 
 
The regularization of $h_{kk}$ is effected by: (i) decomposing the PN-expanded $h_{kk}$ in its various (spin-2) {\it spheroidal harmonics} contributions
 $\propto {}_2S_{lm\omega}$; (ii) transforming (spin-2) spheroidal harmonics into (spin-2) spherical harmonics ${}_2Y_{lm}$ (as an expansion in powers of $\hat a \omega=m \hat a\Omega$); (iii) summing over the \lq\lq magnetic"  number $m$; and, finally, (iv) subtracting the $l\to \infty$ limit of each (PN-expanded) multipolar contribution $h_{kk}^l=\sum_{m=-l}^l h_{kk}^{(lm)}$. We have checked that the regularized value of 
$h_{kk}(r_0)$ is independent of whether $r \to r_0^+$ or $r\to r_0^-$.

Our analytical results are obtained as a double expansion in powers of $y$ (or, alternatively, of $u=y'(y, \hat a)$) and in powers of $\hat a \equiv \hat a_2$. We have pushed the calculation up to $y^{9.5}$ included and $\hat a^4$ included. [Note that $y^{9.5} \sim u^{9.5} \sim \frac{1}{c^{19}}$ correspond to the 8.5 PN level.] When expressing it in terms of $u=y'(y, \hat a)$ (with Eq. \eqref{eq1.27}), our result for $h_{kk}$ reads
\begin{eqnarray}
\label{eq:hkk_gen}
\frac{m_2}{m_1}h_{kk}(u, \hat a) &=& h_{kk}^{\rm Schw}(u)+\hat a h_{kk}^{(1)}(u)+\hat a^2 h_{kk}^{(2)}(u)\nonumber\\
&+&\hat a^3 h_{kk}^{(3)}(u)+\hat a^4 h_{kk}^{(4)}(u)+O(\hat a^5)\,,
\end{eqnarray}
where\footnote{We introduce explicit minus signs on $h_{kk}$ and $\delta U$ because, in view of the mostly-minus signature used in \cite{Keidl:2010pm,Shah:2012gu} that we followed, we actually computed them within the latter signature, while we defined, as most of the literature, including our previous work, $h_{kk}$ and $\delta U \equiv \frac12 h_{kk} / z_1^3$ in the mostly-plus signature.}
\beq
\label{h_kk_schw}
-h_{kk}^{\rm Schw}(u)= +2u -5 u^2 -\frac54 u^3 -\left(-\frac{1261}{24}+\frac{41}{16}\pi^2  \right) u^4+\ldots
\eeq
is the Schwarzschild (non-spinning) result (which is known both numerically \cite{Akcay:2012ea} and to a very high PN order \cite{Bini:2015bla,Kavanagh:2015lva}), and where the spin-dependent contributions read
\begin{widetext}
\begin{eqnarray}
\label{h_kk_various_ord}
-h_{kk}^{(1)}&=&-6 u^{5/2} +9u^{7/2}-\frac{93}{4}u^{9/2}+\left(-\frac{14207}{72}+\frac{241}{48}\pi^2\right)u^{11/2}\nonumber \\
&+&\left(\frac{62041}{384}\pi^2-\frac{20465009}{14400}-\frac{4672}{15}\gamma-\frac{2336}{15}\ln(u)-\frac{1856}{3}\ln(2)\right) u^{13/2}\nonumber\\
&+&\left(-\frac{59993729681}{1411200}+\frac{2185415}{512}\pi^2+\frac{61424}{21}\ln(2)+\frac{12248}{21}\ln(u)+\frac{125168}{105}\gamma-\frac{3888}{7}\ln(3)\right) u^{15/2}\nonumber\\
&-& \frac{686176}{1575}\pi u^8\nonumber\\
&+& \left(-\frac{47144359183457}{457228800}+\frac{6988832}{2835}\gamma-\frac{605312}{567}\ln(2)+\frac{3403696}{2835}\ln(u)+\frac{33534}{7}\ln(3)\right. \nonumber\\
&& \left. -\frac{8776579}{16384}\pi^4+\frac{6377586259}{442368}\pi^2\right) u^{17/2}\nonumber\\
&+& \frac{9093107}{4725}\pi u^9\nonumber\\
&+& \left[-\frac{3268339807}{155925}\ln(u)+\frac{43808}{45}\ln(u)^2-\frac{63064680978612989}{922078080000}+\frac{175232}{45}\gamma\ln(u)\right. \nonumber\\
&& +\frac{12252544}{1575}\ln(2)\ln(u)+\frac{7009035336469}{176947200}\pi^2-\frac{75337409381}{25165824}\pi^4  -\frac{923862722}{22275}\gamma-\frac{19712}{3}\zeta(3)\nonumber\\
&&
-\frac{102426992006}{1819125}\ln(2) +\frac{175232}{45}\gamma^2+\frac{24491392}{1575}\ln(2)^2+\frac{24505088}{1575}\gamma\ln(2) \nonumber\\
&& \left.
-\frac{34368219}{3080}\ln(3)-\frac{1953125}{792}\ln(5)\right]u^{19/2}
+O(u^{10}\ln u)\,,
\end{eqnarray}
\begin{eqnarray}
-h_{kk}^{(2)}&=&  2 u^3+\frac{17}{4} u^5+16 u^4+\left(\frac{2345}{12} -\frac{593}{256}\pi^2\right) u^6 \nonumber\\
&+& \left( \frac{9345583}{14400}-\frac{4493}{384}\pi^2+\frac{528}{5}\gamma +\frac{264}{5}\ln u +208 \ln 2 \right)u^7\nonumber\\
&+&\left(\frac{202703165}{49152}\pi^2-\frac{2030429057}{50400}+\frac{90088}{105}\gamma+\frac{155752}{105}\ln(2)+\frac{45044}{105}\ln(u)+\frac{1458}{7}\ln(3)\right) u^8\nonumber\\
&& +\frac{11128}{105}\pi u^{17/2}\nonumber\\
&&
+\left[
\frac{44891965652561}{619315200}\pi^2-\frac{36383648176111}{50803200}-\frac{9894998}{2835}\gamma-\frac{19929878}{2835}\ln(2)-\frac{3919339}{2835}\ln(u0)+\frac{1536}{5}\zeta(3)\right.\nonumber\\
&&\left.
+\frac{6318}{7}\ln(3)+\frac{417436343}{8388608}\pi^4\right]u^9\nonumber\\
&&
+\frac{60058814}{33075}\pi u^{19/2}
+O(u^{10} \ln u)\,,\nonumber\\
-h_{kk}^{(3)}&=&  -6 u^{9/2}-40 u^{11/2}-\frac{335}{4} u^{13/2} + \left(-\frac{634003}{900}+\frac{115}{48}\pi^2-\frac{192}{5}\ln(2)-\frac{256}{5}\zeta(3)-\frac{192}{5}\ln(u)\right)u^{15/2} \nonumber\\
&+&  \left(-\frac{178438613}{14400}+\frac{3154577}{3072}\pi^2-\frac{3920}{3}\ln(2)-\frac{5552}{15}\ln(u)-\frac{448}{5}\zeta(3)-\frac{9664}{15}\gamma \right)u^{17/2} \nonumber\\
&+&\left(\frac{9800497}{128}\pi^2-\frac{89327249449}{117600}-\frac{370736}{105}\ln(2)-\frac{26688}{35}\zeta(3)-\frac{250928}{105}\gamma-\frac{178712}{105}\ln(u)-\frac{11664}{7}\ln(3)\right) u ^{19/2} \nonumber\\ 
&& +O(u^{10}\ln u)\,,\nonumber\\
-h_{kk}^{(4)}&=& 13 u^6+\frac{381}{4} u^7+ \left(\frac{2203}{6} +\frac{69}{128}\pi^2\right) u^8 \nonumber\\
&& +\left(\frac{22286713}{10080}+\frac{95884607}{3686400}\pi^2-2048\zeta(5)+\frac{1032}{5}\ln(u)-\frac{54784}{23625}\pi^4+\frac{272}{5}\gamma+\frac{34816}{15}\zeta(3)+\frac{1424}{5}\ln(2)\right)u^9\nonumber\\
&& +O(u^{10}\ln u) \,.
\end{eqnarray}
\end{widetext}

Beware that Eq. \eqref{eq:hkk_gen} (with Eqs. \eqref{h_kk_schw} and \eqref{h_kk_various_ord}) yield the functional dependence of $h_{kk}$ on $m_2\Omega$ and $\hat a_2$ through the auxiliary function $u(m_2\Omega, \hat a_2)=y'(y,\hat a)$. Because of the spin-dependence of the relation $y'(y,\hat a)$, the double expansion of $h_{kk}(y,\hat a)$ in powers of $y$ and $\hat a$ would modify the expressions of the coefficients of the various powers of $\hat a$ in Eq.  \eqref{h_kk_various_ord}.

While we were finalizing our calculations, Shah presented, in various conferences \cite{shah_capra2015,shah_MG14}, some 
 analytical results on the PN expansion of the function 
$\delta U(y, \hat a)$. To ease the comparison between ours and his results, let us also present the form that our results take when expressed in terms of
the functional dependence of $\delta U=h_{kk}/(2z_1^3)$ on $y$ (rather than $u=y'(y,\hat a)$) and $\hat a$.
Namely, 
\begin{eqnarray}
\label{deltaUPN}
\frac{m_2}{m_1}\delta U(y, \hat a) &=& \delta U^{\rm Schw}(y)+\hat a \delta U^{(1)}(y)+\hat a^2 \delta U^{(2)}(y)\nonumber\\
&+&\hat a^3\delta U^{(3)}(y)+\hat a^4 \delta U(y)+O(\hat a^5)\,,
\end{eqnarray}
where
\beq
\label{eq2.11}
\delta U^{\rm Schw}(y)=-y -2y^2 -5 y^3 +\left(-\frac{121}{3}+\frac{41}{32}\pi^2  \right) y^4+\ldots
\eeq
is the Schwarzschild result (i.e., $h_{kk}^{\rm Schw}(u)/(2(1-3y)^{3/2})$) and where
\begin{widetext}
{\small
\begin{eqnarray}
\label{eq2.12}
-\delta U^{(1)}&=&
-\frac{7}{3}y^{5/2} -\frac{46}{3}y^{7/2}-77y^{9/2}-\left(\frac{974}{3}+\frac{29}{32}\pi^2\right)y^{11/2}
\nonumber\\
&-& \left( \frac{176}{5} \ln y+ \frac{348047}{150} +\frac{352}{5}\gamma -\frac{6349}{64}\pi^2 +\frac{416}{3}\ln 2\right) y^{13/2} \nonumber\\
&-& \left(\frac{734961481}{22050}-\frac{8911441}{3072}\pi^2+408\gamma+\frac{1052}{5}\ln(y)+\frac{4744}{7}\ln(2)+\frac{972}{7}\ln(3)\right) y^{15/2}\nonumber\\
&-& \frac{33008}{315}\pi y^{8}\nonumber\\
&-& \left(\frac{700704798839}{3572100}-\frac{27925459441}{1327104}\pi^2+\frac{716672}{8505}\gamma+\frac{147872}{1701}\ln(y)+\frac{4440032}{8505}\ln(2)-\frac{162}{7}\ln(3)+\frac{124925059}{393216}\pi^4\right)y^{17/2}
\nonumber\\
&-& \frac{24020077}{66150}\pi y^{9} \nonumber\\
&+&
\left[
-\frac{2167536532386661}{2521307250}+\frac{780002666754601}{7431782400}\pi^2-\frac{40237200436}{16372125}\ln(y)\right.\nonumber\\
&-&
\frac{70924306472}{16372125}\gamma-\frac{144895599176}{16372125}\ln(2)-\frac{43593199495}{16777216}\pi^4+\frac{10841769}{6160}\ln(3)+\frac{2620096}{1575}\ln(2)\ln(y)\nonumber\\
&+&
\frac{437824}{525}\gamma\ln(y)+\frac{5240192}{1575}\gamma\ln(2)-\frac{5504}{5}\zeta(3)-\frac{9765625}{14256}\ln(5)+\frac{109456}{525}\ln(y)^2\nonumber\\
&+&\left.
\frac{437824}{525}\gamma^2+\frac{1744448}{525}\ln(2)^2\right]y^{19/2}\,,
\nonumber\\
-\delta U^{(2)}&=&
y^3+\frac{86}{9} y^4+\frac{577}{9} y^5+\left(\frac{1147}{3}-\frac{593}{512}\pi^2\right) y^6 \nonumber\\
&+&  \left(\frac{1288408}{675}-\frac{92557}{9216}\pi^2+\frac{264}{5}\gamma+104\ln(2)+\frac{132}{5}\ln(y)\right)y^7\nonumber\\
&+& 
\left(\frac{710125279}{294912}\pi^2-\frac{14713942}{945}+\frac{33868}{315}\ln(y)+\frac{6632}{21}\ln(2)+\frac{67736}{315}\gamma+\frac{729}{7}\ln(3)\right) y^8\nonumber\\
&+&  
\frac{5564}{105}\pi y^{17/2}\nonumber\\
&+& 
\left(\frac{597328}{567}\gamma+\frac{7102544}{2835}\ln(2)-\frac{53568695707}{99225}-\frac{486}{7}\ln(3)+\frac{383336}{567}\ln(y)+\frac{71983730742461}{1238630400}\pi^2+\frac{768}{5}\zeta(3)\right.\nonumber\\
&+&\left.
\frac{417436343}{16777216}\pi^4\right)y^9
\nonumber\\
&+&
\frac{10755481}{33075}\pi y^{19/2}\,,
\nonumber\\
-\delta U^{(3)}&=&
-y^{9/2}-\frac{1526}{81} y^{11/2} -\frac{13625}{81} y^{13/2}
-\left(\frac{242891}{225}+\frac{1319}{384}\pi^2+\frac{96}{5}\ln 2+\frac{128}{5}\zeta(3)+\frac{96}{5}\ln y  \right)y^{15/2} \nonumber\\
&-& \left(\frac{13829101}{1215}-\frac{80954347}{165888}\pi^2+\frac{1272}{5}\ln 2+160 \zeta(3)+\frac{2224}{15}\ln y+\frac{1136}{15}\gamma  \right)y^{17/2}\nonumber\\
&-&  
\left(\frac{164644545466}{297675}-\frac{17476082953}{331776}\pi^2+\frac{7481392}{2835}\ln(2)+\frac{3234848}{2835}\gamma+\frac{3560992}{2835}\ln(y)+\frac{7104}{7}\zeta(3)+\frac{1944}{7}\ln(3)\right) y^{19/2}\,,
\nonumber\\
-\delta U^{(4)}&=&
2y^6 +\frac{8120}{243} y^7 +\left(\frac{85420}{243}+\frac{69}{256}\pi^2\right) y^8 \nonumber\\
&+&\left(\frac{884633}{315}+\frac{50786207}{7372800}\pi^2+\frac{232}{5}\ln(2)+\frac{36}{5}\ln(y)+\frac{15488}{15}\zeta(3)-1024\zeta(5)-\frac{27392}{23625}\pi^4+\frac{136}{5}\gamma\right) y^9\,,
\end{eqnarray}
all computed up to  $O(y^{10}\ln y)$.
}
\end{widetext}
The analytical results recently presented by Shah are less accurate than ours (they stop at order $\hat a y^{13/2}$, $\hat a^2 y^6$, $\hat a^3 y^{13/2}$ and $\hat a^4 y^6$, respectively), but agree with ours.

Several features of our results are to be noted:
\begin{enumerate}
  \item The expansion of $h_{kk}$ or $\delta U$ in spin has the structure
\begin{eqnarray}
\label{h_kk_formal}
h_{kk}&\sim& u(1+u+u^2+\ldots)+b(u+u^2+\ldots)\nonumber\\
&+& b^2 (1+u+u^2+\ldots)+b^3 (1+u+u^2+\ldots)\nonumber\\
&+& b^4 (1+u +u^2 +\ldots)
\end{eqnarray}
where $b=\hat a u^{3/2}$ (and where we omitted numerical coefficients in the various PN-correcting parentheses $\varphi(u)=1+u+u^2+\ldots$)
  \item The normal structure of PN expansions [proceeding by successive {\it integer} powers of $u=O(1/c^2)$ in the various correcting parentheses $\varphi(u)=1+u+u^2+\ldots$ entering Eq. \eqref{h_kk_formal} above] breaks down at the {\it fractional} 5.5 PN order i.e., in a term $\propto u^{11/2}$ in \footnote{Note that $c_0=0$ in the term linear in $b$.}
\begin{eqnarray}
\varphi(u) &=& c_0 +c_1 u +c_2u^2 +c_3 u^3 +c_4 u^4 +c_5 u^5 \nonumber\\
&+& c_{5.5}u^{11/2}+c_6 u^6 +c_{6.5}u^{13/2}+\ldots
\end{eqnarray}
[For recent discussions of the similar 5.5PN breakdown of the normal PN expansion in a Schwarzschild background see Refs. \cite{Bini:2013rfa,Blanchet:2013txa}]

 \item It was pointed out in \cite{Damour:2015isa,Bini:2015mza} that the first logarithmic terms in PN expansion (which are linked to the near-zone effect of tails \cite{Blanchet:1987wq,Damour:2009sm,Blanchet:2010zd}) come accompanied by an Euler constant $\gamma$ in the combination $\gamma +\ln (\Omega r_0 /c)$. This \lq\lq Eulerlog" rule is first violated at the 8PN level in nonspinning systems \cite{Bini:2014nfa, Johnson-McDaniel:2015vva}.
By contrast, Eq. (\ref{eq2.12}) shows that, in presence of spin, the Eulerlog rule is violated at the (earlier) 6.5PN level (i.e., in a term $\propto \hat a u^{15/2}$). We shall discuss in future work that the origin of this behavior is linked to the boundary condition at (and the energy flux down) the horizon.
\end{enumerate}

\section{Analytical computation of the self-force corrections to the EOB spin-dependent potentials}

In Eq.~(\ref{12deltaz1}) we have exhibited the connection between $\delta z_1 \equiv - z_1^2 \delta U$ and the GSF corrections $\delta A$, $\delta G_S$ to two of the EOB coupling functions. In order to extract from the {\it single} function of two variables $\delta z_1 (y,\hat a)$ the two separate functions of two variables $\delta A (u,\hat a)$, $\delta G_S (u,\hat a)$, we need to normalize the latter functions by restricting their spin dependence. In the present paper, we use the Damour-Jaranowski-Sch\"afer gauge \cite{Damour:2008qf} where the circular limits of $\delta A$ and $\delta G_S$ are similar to their (zeroth GSF order) Kerr counterparts in that they depend on $u$ but not on $p_{\phi}$. As it is clear from Eq.~(\ref{eq1.20}), $A^{\rm Kerr} (u,\hat a)$ and $G_S^{\rm Kerr} (u,\hat a)$ are {\it even} functions of $\hat a$. It is then natural, and conventionally possible, to restrict the $\hat a$-dependence of $\delta A (u,\hat a)$ and $\delta G_S (u,\hat a)$ by requiring that they are both {\it even} functions of $\hat a$. It is also convenient to: (i) decompose $\delta A (u,\hat a)$ in its spin-independent piece $A_{\rm 1SF}^{(0)} (u)$ and a spin-dependent contribution $\hat a^2 f_A (u,\hat a^2)$; and (ii) introduce some rescaled versions of $f_A$ and $\delta G_S$.

\smallskip

Namely, we write 
\begin{eqnarray}
\label{eq4.1}
\delta A(u,\hat a) &=& A_{\rm 1SF}^{(0)}(u)+\hat a ^2 f_A(u,\hat a^2)\nonumber\\
&=& A_{\rm 1SF}^{(0)}(u)+3u^4 \hat a ^2 f_A^{\rm resc}(u,\hat a^2)\,,\nonumber\\
\delta G_S(u, \hat a)&=& -\frac{5}{8}u^4  \delta G_S^{\rm resc}(u, \hat a^2)\,,
\end{eqnarray}
with the additional decomposition
\begin{eqnarray}
\label{eq4.3}
f_A^{\rm resc}(u,\hat a^2)&=& f_A^{(0)\rm resc}(u)+\hat a^2 f_A^{(2)\rm resc}(u)\nonumber\\
&&+\hat a^4 f_A^{(4)\rm resc}(u)\,,\nonumber\\
\delta G_S^{\rm resc}(u, \hat a^2) &=& \delta G_S^{(0)\rm resc}(u)+\hat a ^2 \delta G_S^{(2)\rm resc}(u)\nonumber\\
&&+\hat a^4\delta G_S^{(4)\rm resc}(u)\,.
\end{eqnarray}
The rescaling factors $3 u^4$ and $-\frac58 u^4$ in Eqs.~(\ref{eq4.1}) are the leading-order (LO) PN contributions so that the PN expansions of both $f_A^{\rm resc}$ and $\delta G_S^{\rm resc}$ start as $1+O(u)$.

\smallskip

We have truncated the latter decompositions in powers of $\hat a^2$ to the $O (\hat a^4)$ level because we shall see below that the truncated expansions (\ref{eq4.3}) allow one to parametrize the numerically known spin dependence of $\delta U(y,\hat a)$ even for large spins $\vert \hat a \vert \leq 0.9$.  Evidently, the exact functions $f_A^{\rm resc} (u , \hat a^2)$ and $\delta G_S^{\rm resc} (u,\hat a^2)$ involve higher powers of $\hat a^2$. On the other hand, our present limited analytical knowledge of $\delta U(y,\hat a)$ allows one to have information about $f_A^{\rm resc} (u , \hat a^2)$ and $\delta G_S^{\rm resc} (u,\hat a^2)$ only through the ${\hat a}^2$ terms.

\smallskip

Identifying the various powers of $u$ and $\hat a$ on both sides of Eq.~(\ref{12deltaz1}) allows  one to convert our analytical results (\ref{eq2.11})--(\ref{eq2.12}) into an analytical knowledge of the PN expansions of $f_A^{(n) {\rm resc}} (u)$ and $\delta G_S^{(n) {\rm resc}} (u)$ (with $n=0,2$), with the following results:

\begin{widetext}
{\small
\begin{eqnarray}
\label{deltaGS_fA_all}
\nu^{-1} \delta G_S^{(0)\rm resc}(u)&=& 1+\frac{102}{5}u +\left(\frac{80399}{720} -\frac{241}{120}\pi^2  \right)u^2 \nonumber\\
&+&  \left(\frac{12015517}{18000}-\frac{62041}{960}\pi^2+\frac{3712}{15}\ln(2)+\frac{9344}{75}\gamma+\frac{4672}{75}\ln u \right)u^3\nonumber\\
&+&\left(-\frac{437083}{256}\pi^2-\frac{122848}{105}\ln(2)+\frac{122119110037}{7056000}+\frac{7776}{35}\ln(3)-\frac{250336}{525}\gamma\right. \nonumber\\
&& \left. -\frac{24496}{105}\ln(u)\right)u^4 \nonumber\\
&+& \frac{1372352}{7875}\pi u^{9/2} \nonumber\\
&+& \left(\frac{753139951463}{17860500}-\frac{6377586259}{1105920}\pi^2+\frac{1210624}{2835}\ln(2)-\frac{67068}{35}\ln(3)-\frac{13977664}{14175}\gamma\right. \nonumber\\
&+& \left. \frac{8776579}{40960}\pi^4-\frac{6807392}{14175}\ln(u)\right)u^5
\nonumber\\
&-&\frac{18186214}{23625}\pi u^{11/2}\nonumber\\
&+& \left[-\frac{350464}{225}\gamma^2+\frac{75337409381}{62914560}\pi^4+\frac{1847725444}{111375}\gamma-\frac{49010176}{7875}\ln(2)\gamma+\frac{34368219}{7700}\ln(3)  \right. \nonumber\\
&&+\frac{556745811545134537}{18441561600000}-\frac{7009035336469}{442368000}\pi^2+\frac{204853984012}{9095625}\ln(2)+\frac{390625}{396}\ln(5)+\frac{39424}{15}\zeta(3)\nonumber\\
&&\left.-\frac{48982784}{7875}\ln(2)^2 +\left(\frac{6536679614}{779625}-\frac{24505088}{7875}\ln(2)-\frac{350464}{225}\gamma\right)\ln(u)-\frac{87616}{225}\ln(u)^2 \right] u^6\nonumber\\
&-& \frac{33472011779}{49116375}\pi u^{13/2} + O(u^7\ln u)\,,
\end{eqnarray}
\begin{eqnarray}
\label{DGS0}
\nu^{-1} \delta G_S^{(2)\rm resc}(u)&=& \frac{16}{5}u +\frac{13}{5} u^2+\frac{181}{5} u^3 \nonumber\\
&+& \left(\frac{1653383}{6000}+\frac{21}{20}\pi^2+\frac{384}{25}\ln(2)+\frac{512}{25}\zeta(3)+\frac{384}{25}\ln(u)\right)u^4\nonumber\\
&+& \left(\frac{3328}{25}\gamma-\frac{870659}{2560}\pi^2+\frac{1376}{5}\ln(2)+\frac{896}{25}\zeta(3)+\frac{28533659}{6000}+\frac{2144}{25}\ln(u)\right)u^5\nonumber\\
&+&\left(\frac{79424}{75}\gamma-\frac{36768409}{1280}\pi^2+\frac{321984}{175}\ln(2)+\frac{15552}{35}\ln(3)+\frac{53376}{175}\zeta(3)+\frac{2032059026897}{7056000}+\frac{127264}{175}\ln(u)\right) u^6 \nonumber\\
&+&   O(u^7\ln u)\,,\nonumber\\
\nu^{-1} f_A^{(0)\rm resc}(u)&=& 1+\frac{11}{4}u+\left(\frac{1769}{72}-\frac{593}{1536}\pi^2\right)u^2
+\left(\frac{1291}{2304}\pi^2+\frac{1752929}{43200}+\frac{104}{3}\ln 2+\frac{88}{5}\gamma+\frac{44}{5}\ln u  \right)u^3 \nonumber\\
&+& \left(-\frac{4401727153}{604800}+\frac{226526909}{294912}\pi^2-\frac{19564}{315}\ln(2)-\frac{4012}{315}\gamma+\frac{243}{7}\ln(3)-\frac{2006}{315}\ln(u)\right)u^4
\nonumber\\
&+& \frac{5564}{315}\pi u^{9/2}\nonumber\\
&+& \left[\frac{24361}{1701}\gamma+\frac{52822399604561}{3715891200}\pi^2+\frac{2473421}{8505}\ln(2)-\frac{891}{7}\ln(3)+\frac{417436343}{50331648}\pi^4+\frac{256}{5}\zeta(3)\right. \nonumber\\
&-& \left. \frac{21340366131467}{152409600}+\frac{1041101}{17010}\ln(u)\right]u^5\nonumber\\
&+& \frac{8414863}{99225}\pi u^{11/2} + O(u^6\ln u)\,,\nonumber\\
\nu^{-1} f_A^{(2)\rm resc}(u)&=& u^2-\frac16 u^3 + \left(\frac{4037}{72}+\frac{23}{256}\pi^2\right)u^4 \nonumber\\ 
&+&
\left(\frac{136}{15}\gamma+\frac{85363007}{22118400}\pi^2+\frac{424}{15}\ln(2)-\frac{27392}{70875}\pi^4+\frac{16256}{45}\zeta(3)+\frac{1447357}{6300}-\frac{1024}{3}\zeta(5)+\frac{76}{5}\ln(u)\right)u^5\nonumber\\
&+&   O(u^6\ln u)\,.
\end{eqnarray}
}
\end{widetext}

The only terms among the above, high-accuracy, PN expansions that were known from standard PN computations were the next-to-leading-order (NLO) corrections to $\delta G_S^{(0){\rm resc}}$ (i.e., $1+\frac{102}5 u$) \cite{Nagar:2011fx,Barausse:2011ys} and to $f_A^{(0) {\rm resc}} (u)$ (i.e., $1+\frac{11}4 u$) \cite{Balmelli:2013zna}.

\section{Numerical computation of the self-force corrections to the EOB spin-dependent potentials}

Shah, Friedman and Keidl \cite{Shah:2012gu} have numerically computed the values of the function $\delta U (y' , \hat a)$ for a sample of radii $\hat r_0 \equiv r_0 / M \equiv 1 / y'\equiv 1/u$ (between 4 and 100) and of dimensionless spin parameters $\hat a$ (namely $\pm \,0.9$, $\pm \,0.7$, $\pm \,0.5$, $0.0$). [Note that the data published in the latter reference  were marred by a technical error; A. Shah kindly provided us with a corrected version of Table III in \cite{Shah:2012gu}; see also Table IV in \cite{vandeMeent:2015lxa}.] For a sub-sample of the latter data, Shah et al. computed $\delta U (y',\hat a)$ for various values of $\hat a$ but for the same value of the inverse radius\footnote{Note that Ref.~\cite{Shah:2012gu} parametrizes the circular orbits by the Boyer-Lindquist radius $\hat r_0$, which corresponds to fixing the modified frequency parameter $y'$, Eq.~(\ref{eq1.27}), such that $u = 1/\hat r_0 = y'(y, \hat a)$.}. In view of the (assumed) even spin dependence of $\delta A$ and $\delta G_S$ in Eq.~(\ref{12deltaz1}), we can then extract the numerical values of $\delta A(u,\vert \hat a \vert)$ and $\delta G_S (u,\vert \hat a \vert)$ by suitably projecting both sides of Eq.~(\ref{12deltaz1}) (after multiplying them by appropriate, known Kerr factors). Explicitly we find
\beq
\label{eq5.1}
\delta G_S(u,\hat a) =\frac{1}{\hat a }\,  \frac{\left[z_1^K\left(\frac12 \delta z_1-{\mathcal K}  \right)\right]^{\rm odd}}{[p_\phi^K z_1^K]^{\rm even}}\,,
\eeq
\beq
\label{eq5.2}
\delta A(u, \hat a) =2 \frac{\left[\frac{z_1^K}{p_\phi^K}\left(\frac12 \delta z_1-{\mathcal K}  \right)\right]^{\rm even}}{\left[\frac{1}{ p_\phi^K z_1^K }\right]^{\rm even}} \,.
\eeq
Here, both sides are to be evaluated at the same value of $u=y' =u^{\rm circ} (y)$, the superscripts $K$ denote taking a Kerr (-circular) value, and the superscripts ``odd'' or ``even'' denote the operation of taking, respectively, the odd or even part of a function of $\hat a$, namely
\begin{eqnarray}
[F (u,\hat a)]^{\rm even} &\equiv& \frac12 (F(u,\hat a) + F(u,-\hat a))\nonumber\\
{}[F (u,\hat a)]^{\rm odd}  &\equiv& \frac12 (F(u,\hat a) - F(u,-\hat a))\,.
\end{eqnarray}
The numerical values of the {\it rescaled} functions $\delta G_S^{\rm resc} (u, \vert \hat a \vert)$ and $f_A^{\rm resc} (u, \vert \hat a \vert)$ we found by applying the above projection formulas to the corrected data communicated by Shah are listed in Tables \ref{tab:1} and \ref{tab:2} below.

\smallskip

  
\begin{table}
\centering
\caption{Numerical values for $\delta G_S^{\rm resc}(u, \vert\hat a\vert)$.}
\begin{ruledtabular}
\begin{tabular}{cccc}
$u$ & $\delta G_S^{\rm resc}(u, 0.5)$ & $\delta G_S^{\rm resc}(u, 0.7)$   & $\delta G_S^{\rm resc}(u, 0.9)$  \cr
\hline
 1/100 & 1.221(5)         &    1.229(4)  &1.239(3) \cr
 1/70  & 1.322(2)     &    1.333(1)      & 1.3478(8)   \cr
 1/50  & 1.4614287(1)   &    1.47711466(7)    & 1.49802501(5) \cr
 1/30  & 1.81295880(4)    &   1.83965095(4)     &1.87522164(2) \cr
 1/20  & 2.30788136(4)    &   2.34941806(1)      &  2.40474331(2)\cr
 1/15  & 2.8736(2)      &   2.93195(2)         & 3.009697(3) \cr
 1/10  & 4.27400044(4)    &   4.37851888(3)      &4.51778792(8) \cr
 1/8   & 5.64616283(1)                 &-                     &-\cr
\end{tabular}
\end{ruledtabular}
\label{tab:1}
\end{table}

\begin{table}
\centering
\caption{Numerical values for $f_A^{\rm resc}(u, \vert\hat a\vert)$.}
\begin{ruledtabular}
\begin{tabular}{cccc}
$u$ & $f_A^{\rm resc}(u, 0.5)$         & $f_A^{\rm resc}(u, 0.7)$   & $f_A^{\rm resc}(u, 0.9)$  \cr
\hline
 1/100 &  1.0(5)                          &   1.0(2)         & 1.0(1)   \cr
 1/70  &  1.04(9)                         &   1.04(5)        & 1.04(3)   \cr
 1/50  &  1.063828(5)                     &   1.063926(2)    & 1.064057(2)   \cr
 1/30  &  1.1173441(9)                    &   1.1176295(8)   & 1.1180100(3)   \cr
 1/20  &  1.1993228(7)                    &   1.2000319(4)   & 1.2009773(4)   \cr
 1/15  &  1.302(3)                        &   1.3037(2)      & 1.30564(2)   \cr
 1/10  &  1.5991204(3)                    &   1.6042429(2)   & 1.6110832(4)    \cr
 1/8   &  1.9497140(1)                    &    -             & - \cr
\end{tabular}
\end{ruledtabular}
\label{tab:2}
\end{table}

In these Tables the digits within parentheses indicate a rough estimate of the numerical uncertainty on the last digit of the corresponding numerical values of $\delta G_S^{\rm resc}$ and $f_A^{\rm resc}$. These error estimates were obtained from the error estimates on $\delta U (y' , \hat a)$ kindly communicated by A. Shah.

\smallskip

By looking at the $\hat a$-dependence of $\delta G_S^{\rm resc}$ and $f_A^{\rm resc}$ (for a fixed value of $u=y'$) in Tables~\ref{tab:1} and \ref{tab:2} one sees immediately that it is rather mild. We found that it is numerically accurate to use the truncated expansions (\ref{eq4.3}) to fit the numerical data. [Note that these expansions include one more power of ${\hat a}^2$ than the ones we could extract from our analytical results.]  For each function $\delta G_S^{\rm resc}$ or $f_A^{\rm resc}$, and for each value of $u$ in Tables~\ref{tab:1} and \ref{tab:2}, we can use the three numerical results for $\hat a^2 = (0.5)^2$, $\hat a^2 = (0.7)^2$ and $\hat a^2 = (0.9)^2$ to extract (by solving a linear system of three equations in three unknowns) the corresponding numerical values of the coefficient functions $\delta G_S^{(n){\rm resc}} (u)$, $f_A^{(n) {\rm resc}} (u)$ (with $n = 0, 2$ and $4$). The numerical values we found for these coefficient functions are listed in Tables~\ref{tab:3} and \ref{tab:4} below. We did not propagate the numerical errors affecting $\delta G_S^{\rm resc}$ and $f_A^{\rm resc}$ to their corresponding coefficient functions $\delta G_S^{(n){\rm resc}} (u)$, $f_A^{(n) {\rm resc}} (u)$.

\begin{widetext}

\begin{table}
\centering
\caption{Numerical values for $\delta G_S^{(n)\rm resc}(u)$.}
\begin{ruledtabular}
\begin{tabular}{cccccc}
$u$ & $\delta G_S^{(0)\rm resc}(u)$         & $\delta G_S^{(2)\rm resc}(u)$   & $\delta G_S^{(4)\rm resc}(u)$  \cr
\hline
 1/100 &    1.21297347                  &  0.32298056 $\times 10^{-1}$      &  $-$0.23967471   $\times 10^{-5}$   \cr
 1/70  &    1.31023001                  &  0.46361889 $\times 10^{-1}$      &  $-$0.88948253   $\times 10^{-5}$    \cr
 1/50  &    1.44508613                  &  0.65376139 $\times 10^{-1}$      &  $-$0.24062736   $\times 10^{-4}$    \cr
 1/30  &    1.78514161                  &  0.11129505                     &  $-$0.10509689   $\times 10^{-3}$    \cr
 1/20  &    2.26457500                  &  0.17330498                     &  $-$0.31813647   $\times 10^{-3}$     \cr
 1/15  &    2.81269577                  &  0.24365352                     &  $-$0.54588538   $\times 10^{-3}$    \cr
 1/10  &    4.16506631                  &  0.43586052                     &  $-$0.49597783   $\times 10^{-3}$    \cr 
\end{tabular}
\end{ruledtabular}
\label{tab:3}
\end{table}

\begin{table}
\centering
\caption{Numerical values for $f_A^{(n)\rm resc}(u)$.}
\begin{ruledtabular}
\begin{tabular}{cccccc}
$u$ & $f_A^{(0)\rm resc}(u)$         & $f_A^{(2)\rm resc}(u)$   & $f_A^{(4)\rm resc}(u)$  \cr
\hline
 1/100 &       1.03848559            &  $-$0.19044446  $\times 10^{-1}$     &   0.12349670 $\times 10^{-1}$   \cr
 1/70  &       1.04376926            &  $-$0.24085514  $\times 10^{-4}$      &   0.14850697 $\times 10^{-3}$    \cr
 1/50  &       1.06372483            &   0.41107182  $\times 10^{-3}$     & $-$0.14807331 $\times 10^{-5}$    \cr
 1/30  &       1.11704682            &   0.11890403  $\times 10^{-2}$      &   8.51657317 $\times 10^{-8}$    \cr
 1/20  &       1.19858425            &   0.29543259  $\times 10^{-2}$      &   1.22022631 $\times 10^{-7}$     \cr
 1/15  &       1.30055059            &   0.66125741  $\times 10^{-2}$      & $-$0.40611727 $\times 10^{-3}$    \cr
 1/10  &       1.59379299            &   0.21293650  $\times 10^{-1}$      &   0.64609073 $\times 10^{-4} $   \cr
\end{tabular}
\end{ruledtabular}
\label{tab:4}
\end{table}

\begin{table}
\centering
\caption{Fractional errors for  $\delta G_S^{(0){\rm resc}}(u)$, $\delta G_S^{(2){\rm resc}}(u)$, $f_A^{(0){\rm resc}}(u)$, $f_A^{(2){\rm resc}}(u)$.
}
\begin{ruledtabular}  
\begin{tabular}{ccccc}
$u$ & ${\rm  FE } (\delta G_S^{(0){\rm resc}}(u))$& ${\rm FE}(\delta G_S^{(2){\rm resc}}(u))$ & ${\rm FE}(f_A^{(0){\rm resc}}(u))$ & ${\rm FE}(f_A^{(2){\rm resc}}(u))$ \cr
\hline                                                                     
 1/100    &0.16914269 $\times 10^{-3}$ &   0.24269950   $\times 10^{-4}$ & $-$0.85379705$\times 10^{-2}$  & $-$0.10052734  $\times 10^{-1}$\cr  
 1/70  &$-$0.14864240 $\times 10^{-5}$ &   0.93265904   $\times 10^{-6}$ & $-$0.10327051$\times 10^{-3}$  & $-$0.95582144$\times 10^{-1}$\cr   
 1/50     &3.09293406 $\times 10^{-8}$ & $-$0.12947379  $\times 10^{-5}$ &    0.74014351 $\times 10^{-6}$ &$-$0.58614196$\times 10^{-2}$\cr   
 1/30     &5.12276959 $\times 10^{-7}$ & $-$0.16597436  $\times 10^{-4}$ & $-$0.65484497 $\times 10^{-5}$ & $-$0.17784307$\times 10^{-2}$\cr  
 1/20     &0.55605811 $\times 10^{-5}$ & $-$0.15761481  $\times 10^{-3}$ &  $-$0.73973727 $\times 10^{-4}$&$-$0.96906596$\times 10^{-2}$\cr  
 1/15     &0.16269266 $\times 10^{-4}$ & $-$0.44726979  $\times 10^{-3}$ & $-$0.29222881 $\times 10^{-3}$ &$-$0.10663031\cr  
 1/10     &0.16054604 $\times 10^{-3}$ & $-$0.10111086  $\times 10^{-1}$ &$-$0.45501398$\times 10^{-2}$   &$-$0.12969213\cr  
\end{tabular}
\end{ruledtabular}
\label{tab:5}
\end{table} 

\begin{table}
\centering
\caption{Numerical vs theoretical  values for $\delta U(u, \hat a=0.5)$.
}
\begin{ruledtabular}  
\begin{tabular}{cccc}
$u$ &$\delta U^{\rm num} (u, \hat a=0.5)$ (Shah \cite{priv_com}) &$\delta U^{\rm PN\mbox{-}theor}/ \delta U^{\rm num}-1$   & $\delta U^{\rm EOB\mbox{-}theor}/ \delta U^{\rm num}-1$   \cr
\hline                                                                     
 1/100 &$-$0.0101896245(5)  &$-$0.00000001(5)  &    $-$0.00000001(5)  \cr
 1/70  &$-$0.0146705787(5) &$-$0.00000000(3)    &   $-$0.00000000(3)  \cr
 1/50  &$-$0.02075117876615(8) &$-$0.000000000006(4)   & $-$0.00000000053(4) \cr
 1/30  &$-$0.0354163457476(3) &$-$0.000000000941(4)   & $-$0.00000000203(4) \cr
 1/20  &$-$0.054722233077(1) &$-$0.0000000327(2) & $-$0.00000002109 (2)  \cr
 1/15  &$-$0.07519039(3) &$-$0.0000004(4)     & $-$0.0000002(4)   \cr
 1/10  &$-$0.12016503504(2) &$-$0.0000123021(2) &  $-$0.00000724874(2)   \cr
1/8    &$-$0.15830027075(1)  &$-$0.00008135993(6)  & $-$0.00005156354(6) \cr
1/7    &$-$0.18858304135(2) &$-$0.0002521042(1) &   $-$0.0001621950(1) \cr
1/6    &$-$0.2342693592(1) &$-$0.0009347921(5) &   $-$0.0005626589(4)  \cr
1/5    &$-$0.3135069374(1) &$-$0.0044620321(3)  &  $-$0.0016218169(3)  \cr 
\end{tabular}
\end{ruledtabular}
\label{tab:6}
\end{table} 

\end{widetext}

\section{Comparison between analytical and numerical results}

\begin{figure*}
\begin{center}
\[
\begin{array}{cc}
\includegraphics[scale=0.4]{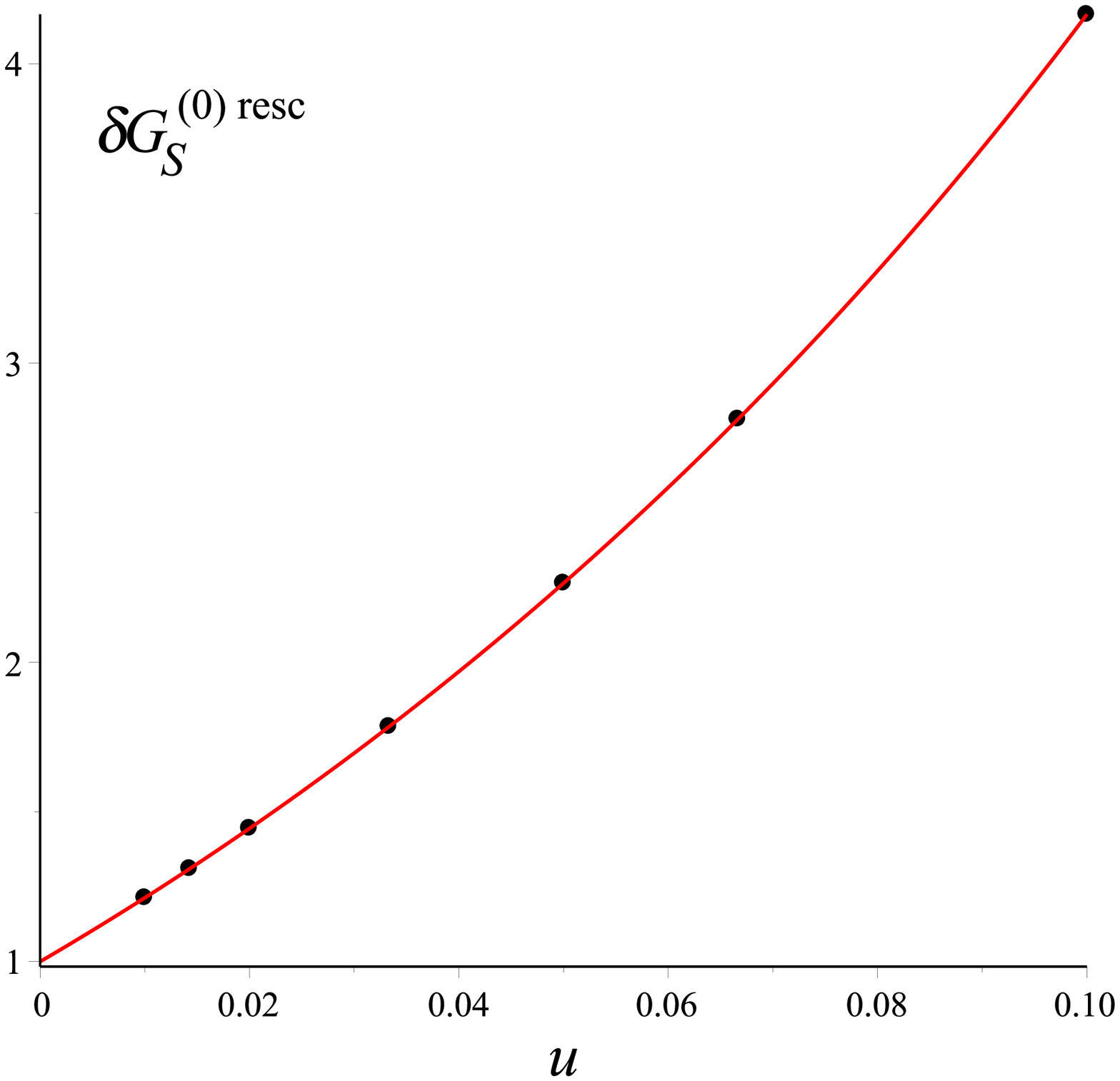} & \includegraphics[scale=0.4]{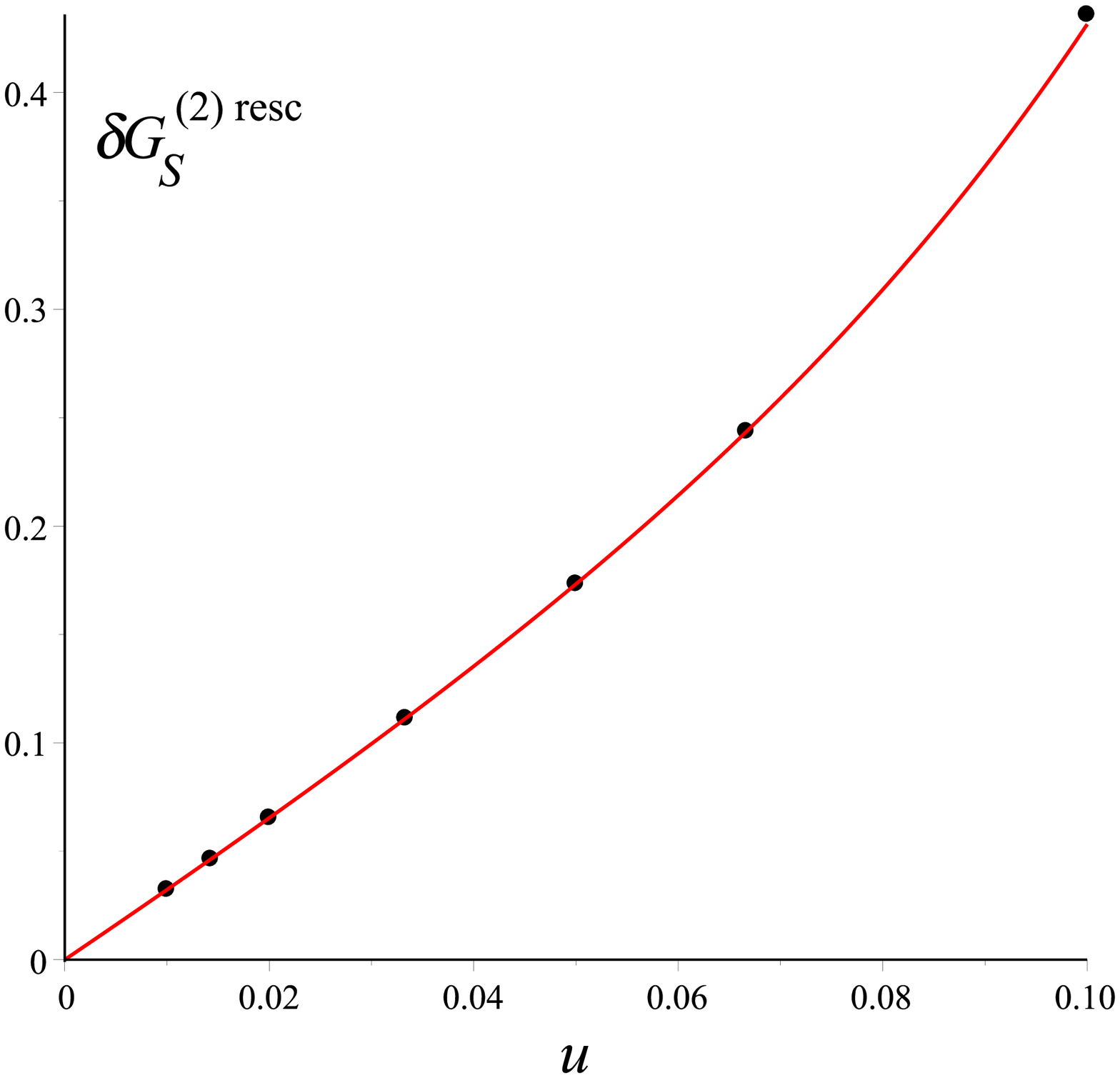} \cr
(a) & (b) \cr
\end{array}
\]
\end{center}
\caption{\label{fig:2}Panel (a): The data points and the theoretical predictions for $\delta G_S^{(0)\rm resc}(u)$.
Panel (b):  The data points and the theoretical predictions for $\delta G_S^{(2)\rm resc}(u)$.
}
\end{figure*}

\begin{figure*}
\begin{center}
\[
\begin{array}{cc}
\includegraphics[scale=0.4]{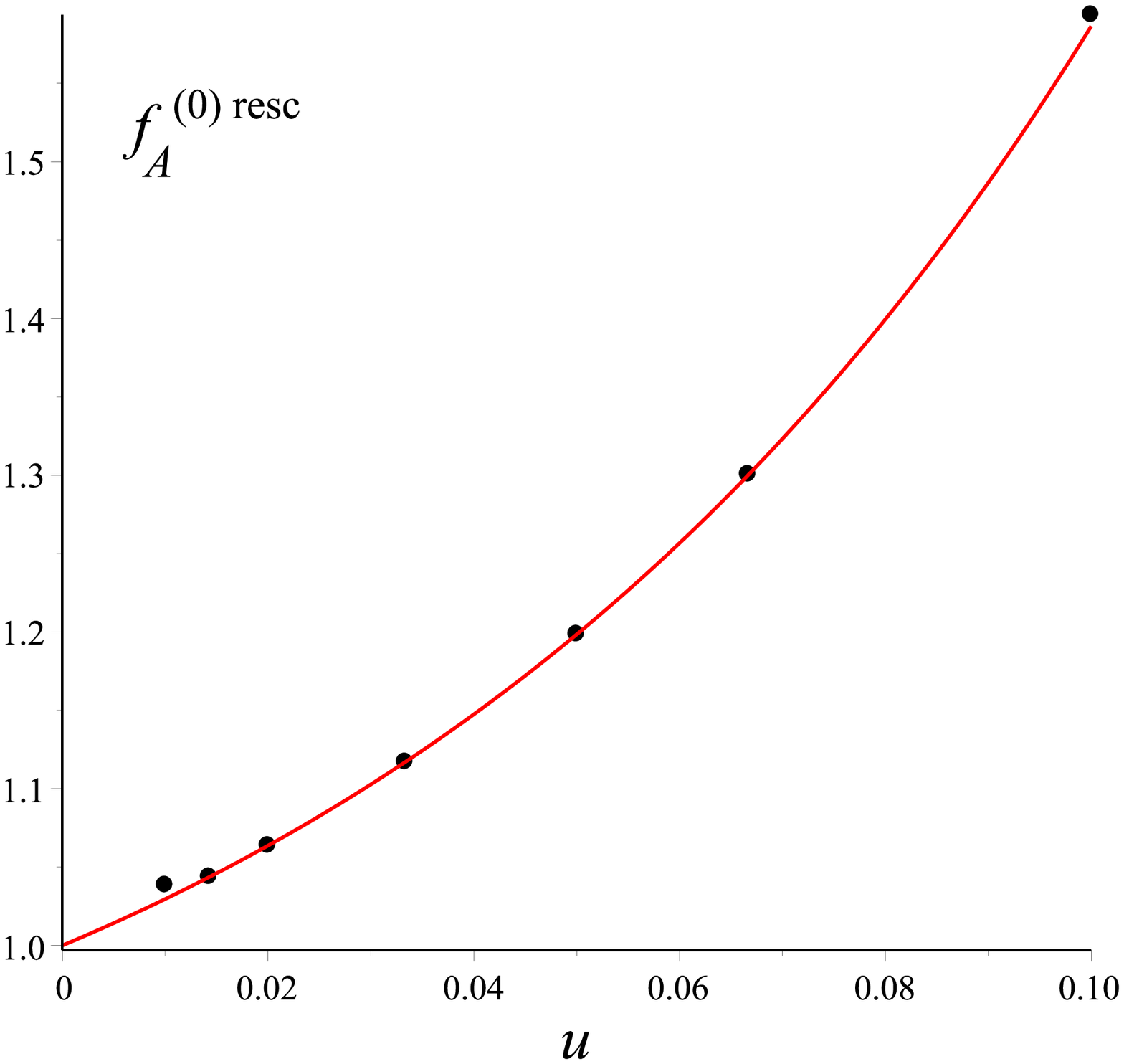} & \includegraphics[scale=0.4]{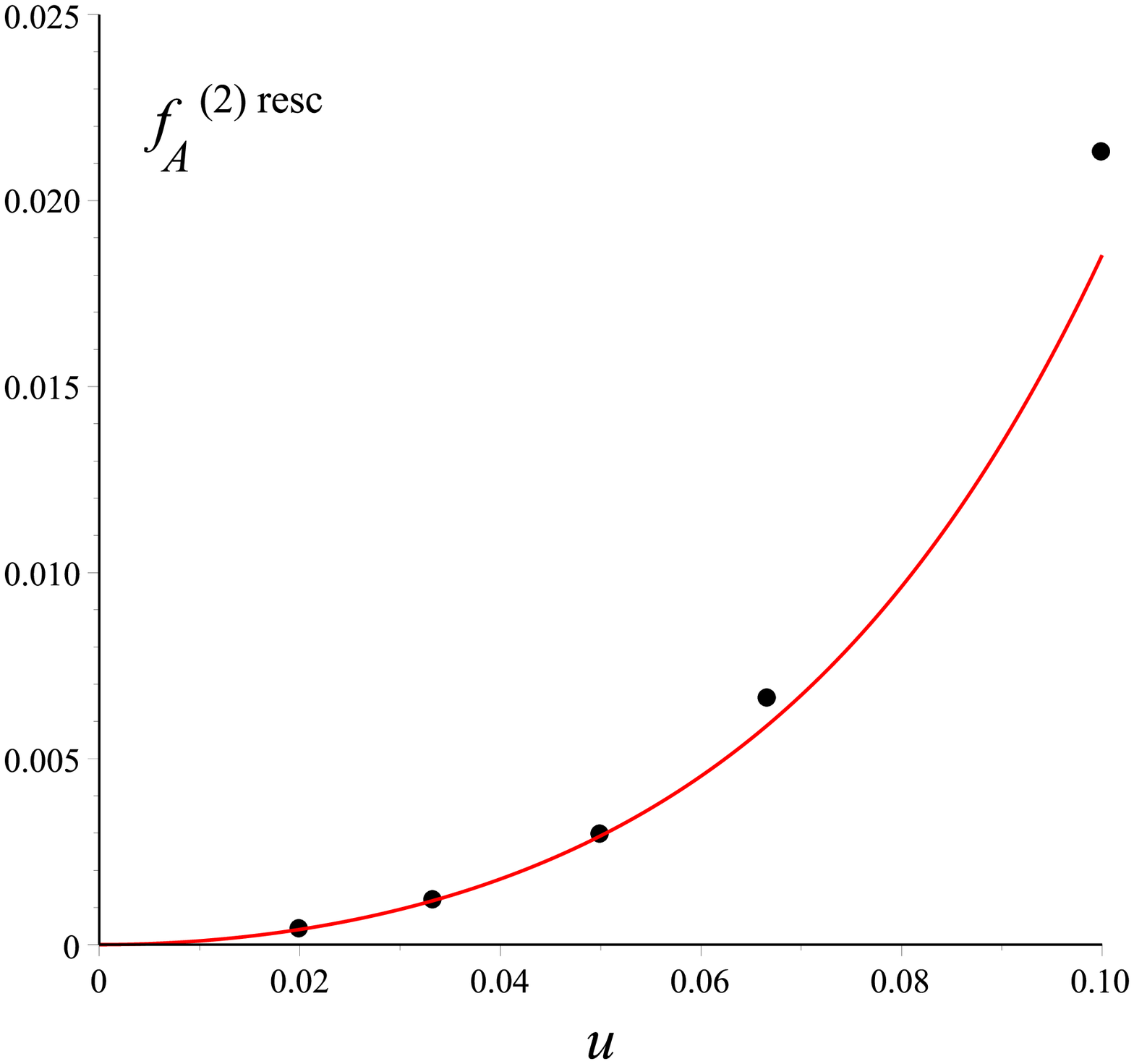} \cr
(a) & (b) \cr
\end{array}
\]
\end{center}
\caption{\label{fig:1}Panel (a): The data points and the theoretical predictions for $f_A^{(0)\rm resc}(u)$.
Panel (b):  The data points and the theoretical predictions for $f_A^{(2)\rm resc}(u)$.
}
\end{figure*}

In Figs.~\ref{fig:2} and \ref{fig:1} we compare our current analytical expressions of the various EOB potentials $\delta G_S^{(n) {\rm resc}} (u)$, $f_A^{(n) {\rm resc}} (u)$ to their numerical counterparts, extracted from the strong-field data computed by Shah et al. As we see, there is an excellent visual agreement between the numerical results (indicated by discrete dots) and the analytical ones (continuous curves) for $\delta G_S^{(0) {\rm resc}} (u)$, $\delta G_S^{(2) {\rm resc}} (u)$ and $f_A^{(0) {\rm resc}} (u)$.  The only function for which there are noticeable differences is $f_A^{(2) {\rm resc}} (u)$. 
The corresponding fractional errors (defined as ${\rm FE}(X)\equiv X^{\rm theor}/X^{\rm num}-1$)
are displayed in Table \ref{tab:5}.

It is then possible to improve the numerical/analytical agreement by adding some (effective) higher-order contributions to $f_A^{(2) {\rm resc}} (u)$, say
$f_A^{(2)\rm  resc, fit}(u)=f_A^{(2)\rm  resc, PN}(u)+(c_1+ c_2\ln (u))u^6$.
 By fitting the numerical-minus-analytical difference we found the following estimate of the higher-order coefficients:  $c_1=-24303.04$ and $c_2=-11754.74$.
We have instead no  analytical  prediction for both $\delta G_S^{(4)\rm  resc}$ and $f_A^{(4)\rm  resc}$, which would need  an analytical knowledge of $\delta U$ at higher orders in $\hat a$. On the other hand, we found that  the data points for the rescaled quantity $u^{-3}\delta G_S^{(4)\rm  resc}$ can be easily fitted. For example, a quadratic fit of the form $262.1643 u^2+0.2878 u -3.1256$ shows a reasonable  agreement  with the existing data points. We refrained from similarly fitting for higher-order corrections to $\delta G_S^{(0)\rm  resc}$, $\delta G_S^{(2)\rm  resc}$ and $f_A^{(0)\rm  resc}$, except for $\delta G_S^{(0)\rm  resc}$ for which we found a good fit (within  $5.6\times 10^{-6}$) of numerical data to $\delta G_S^{(0)\rm  resc \, fit} = \delta G_S^{(0)\rm  resc \, PN} +(983.35+3330.99 \ln(u))u^7$. The data points for  $f_A^{(4)\rm  resc}$ are affected by large errors and a fit in this case does not seem meaningful.

\smallskip

Because of the need to have in hands numerical data with the same value of $u$ and several pairs of opposite values of $\hat a$, the numerical values of the various extracted EOB potentials in Tables \ref{tab:3} and \ref{tab:4} were limited to the semi-strong-field region $0 < u \leq 0.1$. In order to gauge the validity of our analytical results for larger values of $u$, we compared the values of the redshift correction $\delta U (u,\hat a)$ predicted by inserting in the rhs of Eq.~(\ref{12deltaz1}) our analytical PN-expanded results \eqref{eq2.12} to the (corrected) numerical data of Ref.~\cite{Shah:2012gu} (using $\delta z_1^{\rm num} \equiv -z_1^2 \delta U^{\rm num}$). 

We display in Table \ref{tab:6} the ratios $\delta U^{\rm analytical} / \delta U^{\rm numerical} - 1$ for two different analytical predictions of $\delta U$: either the straightforward PN expansion \eqref{deltaUPN}  or its EOB-theoretical form \eqref{12deltaz1} (in which we use the PN expansions of the functions $\delta G_S^{(n)}$ and $\delta f_A^{(n)}$, Eqs.  \eqref{deltaGS_fA_all}, and  the numerical knowledge of $A_{\rm 1SF}^{(0)}(u) $ as given by model 14 in Ref. \cite{Akcay:2012ea}).
For information, we indicate an estimate of the fractional numerical uncertainty on $\delta U^{\rm numerical}$ communicated by Shah. Let us first note that the EOB version of our analytical estimate is systematically more accurate than the corresponding PN estimate. The analytical/numerical agreement is (as expected) excellent in the weak-field regime~\footnote{However, the rather large errors on the data points at $\hat r_0 = 100$ and $70$ show that these points do not bring meaningful information beyond our analytical results.} ($u \ll 1$) and stays rather good (especially for the EOB version) in the strong-field regime (see the $u = \frac15$ EOB data point which agrees within $1.6 \times 10^{-3}$ with the numerical data).

\smallskip

We leave to future work a study of methods for improving the analytical/numerical agreement. In particular, we know from the arguments of Ref.~\cite{Akcay:2012ea} that $\delta z_1$ will blow up proportionally to $1/z_1^2$ near the light ring (where $z_1 \to 0$) or, equivalently, that $\delta U$ will blow up proportionally to $U^4 = 1/z_1^4$ there. As explained in Ref.~\cite{Akcay:2012ea}, this blow up suggests that one should introduce in the concerned EOB potentials some $p_{\phi}$ dependence. However, the introduction of such a $p_{\phi}$ dependence will, in turn, modify the parity of the functional dependence on $\hat a$ of the concerned EOB potentials. [Indeed, we see on Eq.~\eqref{Kerrpphi} that the circular value of $p_{\phi}$ has no well-defined parity in $\hat a$.] Let us finally mention that, in order to achieve a more complete knowledge of $f_A$ and $\delta G_S$ in the strong-field domain, it would be necessary to have more numerical data on $\delta U$, with some suitably chosen sampling of the $(u,\hat a)$ plane.
In particular, data for small values of $\hat a$ would be useful for controlling the strong-field behavior of $\delta G_{S}^{(0)}(u)$ which is of most physical interest (see end of Conclusions).

\begin{figure}
\begin{center}
\includegraphics[scale=0.4]{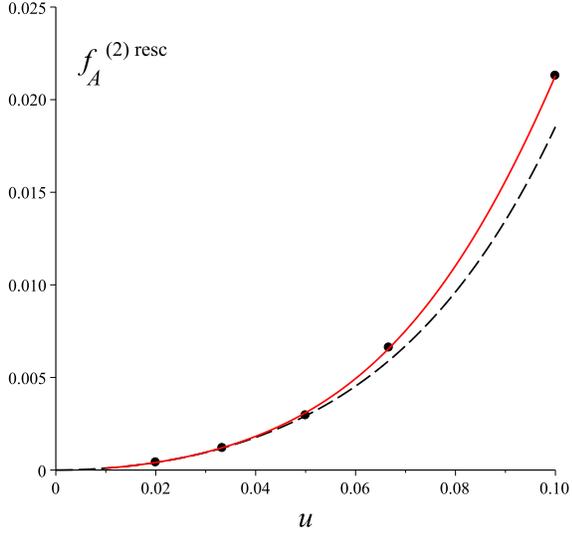} 
\end{center}
\caption{\label{fig:3} The data points for  $f_A^{(2)\rm  resc}$, the theoretical PN prediction (dashed curve) and the fit $f_A^{(2)\rm  resc, fit}(u)=f_A^{(2)\rm  resc, PN}(u)+(c_1+ c_2\ln (u))u^6$ where
$c_1=-24303.04$ and $c_2=-11754.74$.[The  accuracy of the fit is found to be $2.30 \times 10^{-4}$.] 
}
\end{figure}

\begin{figure}
\begin{center}
\includegraphics[scale=0.4]{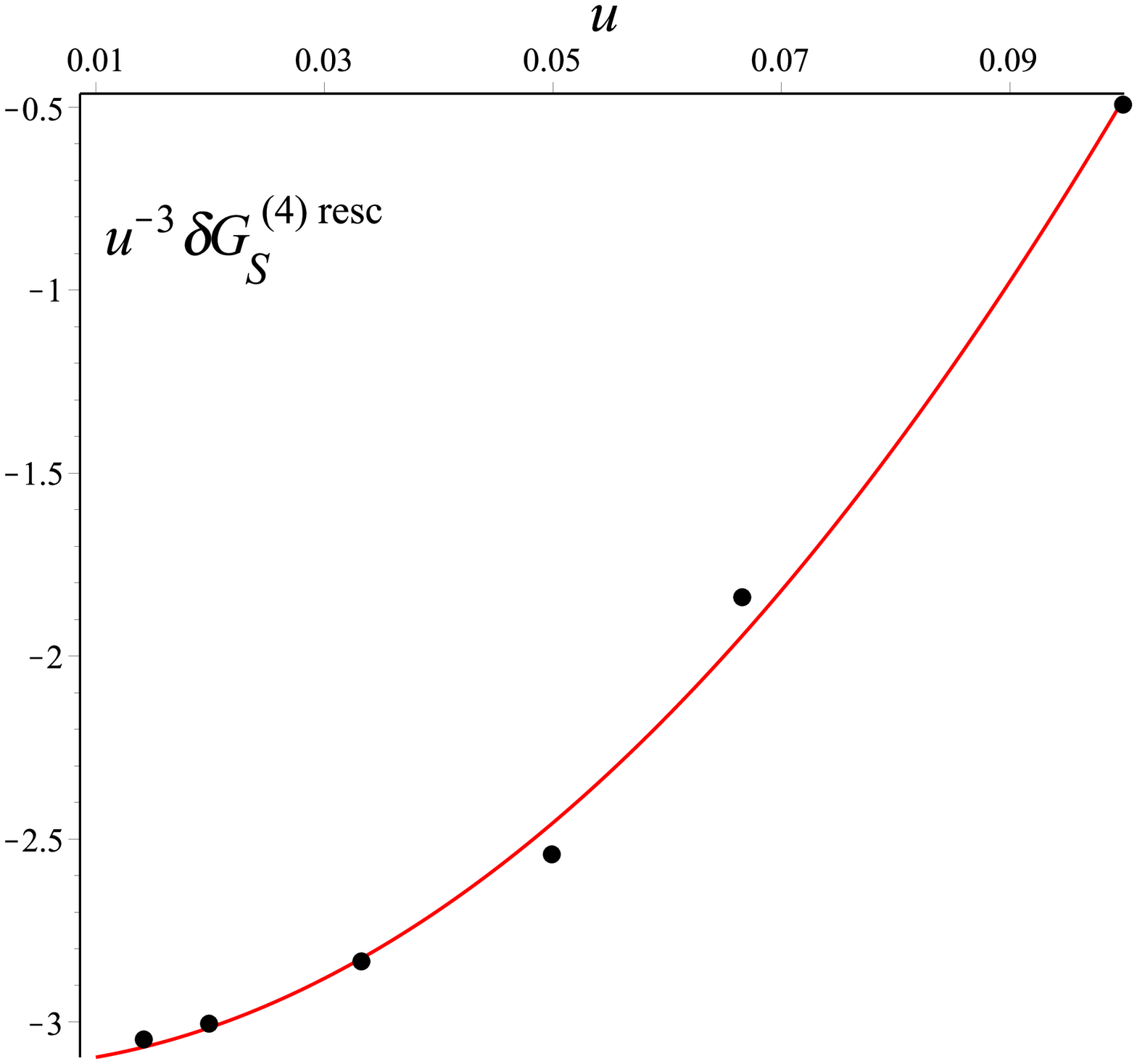} 
\end{center}
\caption{\label{fig:4} The data points for the rescaled quantity $\frac{1}{u^3}\delta G_S^{(4)\rm resc}(u)$ for which   no theoretical prediction is available.
The solid curve superposed to the points corresponds to a (quadratic) fit by the function
$262.1643 u^2+0.2878 u -3.1256$.
}
\end{figure}

\section{Conclusions}

Let us summarize our main results:

\smallskip

We derived the very simple relation Eq.~(\ref{link}) between the GSF correction $\delta z_1$ to the redshift (considered as a function of the orbital frequency) and the $O(m_1^2)$ contribution to the two-body Hamiltonian (considered as a function of phase space variables). The latter relation then implied the simple relation (\ref{12deltaz1}) between $\delta z_1$ and the $O(\nu)$ contributions to the EOB coupling functions $A$ and $G_S$.

\smallskip

We analytically computed the PN expansion of $\delta z_1$ (or, equivalently, $\delta U = -\delta z_1 / z_1^2$) up to order $O(u^{9.5})$ included and $O(\hat a^4)$ included. See Eqs.~\eqref{eq2.12}. We then converted the latter expansions (using Eq.~(\ref{12deltaz1})) into correspondingly accurate PN expansions of the $O(\nu)$ corrections $\delta A$, $\delta G_S$ to the EOB coupling function $A , G_S$. The latter results represent drastic improvements in our knowledge of the spin-dependent interactions encoded within the EOB potentials $A$ and $G_S$.

\smallskip

Going beyond PN expansions (whose validity is a priori limited to the weak-field domain $u \ll 1$), we showed how to extract the numerical values of $\delta A$ and $\delta G_S$ in the strong-field domain $u=O(1)$ from the numerical GSF calculations of $\delta z_1$ \cite{Shah:2012gu}. See Eqs.~(\ref{eq5.1}), (\ref{eq5.2}) and Tables~\ref{tab:1} and \ref{tab:2}. We then compared the latter numerical results to our high-accuracy PN expansions and found excellent agreement when $u \ll 1$, and a good agreement ($\sim 10^{-3}$) up to $u=0.20$ (corresponding to $r_0 = 5M$).

\smallskip

Let us finally discuss what is probably the physically most important result of the present work. It concerns the main EOB spin-orbit coupling function $G_S$. Both our analytical results and our GSF-extracted numerical data show that the {\it rescaled} GSF correction $\delta G_S^{\rm resc} (u,\hat a)$ significantly increases\footnote{Similarly to a corresponding increase of $\delta A$ when $\hat a = 0$ \cite{Akcay:2012ea}, this increase is linked to the blow up of $\delta G_S$ at the light-ring.} from its value $1$ at large separations $(u \to 0)$ to values of order $5$ at separations of order $r_0 \simeq 10M$ (i.e. $u \simeq 0.1$). However, the LO rescaling factor used for $\delta G_S$ is {\it negative}, and equal to $-\frac58 u^4$. This means that the GSF correction tends to {\it diminish} the value of the total spin-orbit coupling. This confirms what was found in the previous (less accurate) PN calculations \cite{Damour:2008qf,Nagar:2011fx,Barausse:2011ys}.
Let us consider for simplicity the $\hat a \to 0$ limit of $G_S$ and work with the Kerr-rescaled spin-orbit coupling
\begin{eqnarray}
\hat G_S (u,\nu , \hat a_1 , \hat a_2) &=& \frac{G_S (r,m_1 , m_2 , S_1 , S_2)}{G_S^{\rm Kerr} (r, m_1 + m_2 , a_1 + a_2)} \nonumber\\
&=& 1 + \frac{\delta G_S}{G_S^{\rm Kerr}}
\end{eqnarray}
(taken for $S_1 = 0 = S_2$). The current (combined PN and GSF) knowledge of the latter function is
\begin{eqnarray}
\label{GSnu}
\hat G_S (u,\nu ,0,0) &= &1 - \frac5{16}  u \delta G_S^{(0){\rm resc}} (u) -\frac1{16} \nu^2 u^2 \varphi_1 (u)\nonumber\\
&+ &O(\nu^3) \\
&= &\left( 1 + \frac5{16}  u \delta G_S^{(0){\rm resc} } (u) \right. \nonumber\\
&&\left. + \frac{41}{256} \nu^2 u^2 \varphi_2 (u)  
 +O(\nu^3) \right)^{-1} 
\,, \nonumber
\end{eqnarray}
where $\varphi_n (u) = 1+O(u)$ denotes a generic PN correction factor. In the second equation we have used an inverse resummation of   $\hat G_S$
as found useful in recent EOB work.
Damour and Nagar \cite{Damour:2014sva} have a provided an effective expression for $1/ \hat G_S (u,\nu ,0,0)$, parametrized by a constant $c_3$ as indicated below
\begin{eqnarray}
\label{eqDN}
[\hat G_S (u,\nu ,0,0)]^{-1} &=&  1+\frac{5}{16}\nu u \left(1+\frac{102}{5}u \right)\nonumber\\
&& +c_3\nu u^3+\frac{41}{256}\nu^2 u^2\,.
\end{eqnarray}
Using our result, Eq.\eqref{deltaGS_fA_all}, for $\delta G_S^{(0){\rm resc}} (u)$ we can define an effective function of $u$, $c_3^{\rm eff}=c_3^{\rm eff}(u)$, such that the
replacement $c_3 \to c_3^{\rm eff}(u)$ in \eqref{eqDN} is consistent with  our full PN-expanded result.
The value at $u=0$ of this $c_3^{\rm eff}(u)$ is found to be
\beq
c_3^{\rm eff}(0) =\frac{80399}{2304}-\frac{241}{384}\pi^2 \approx 28.7012\,.
\eeq

\begin{figure}
\begin{center}
\includegraphics[scale=0.4]{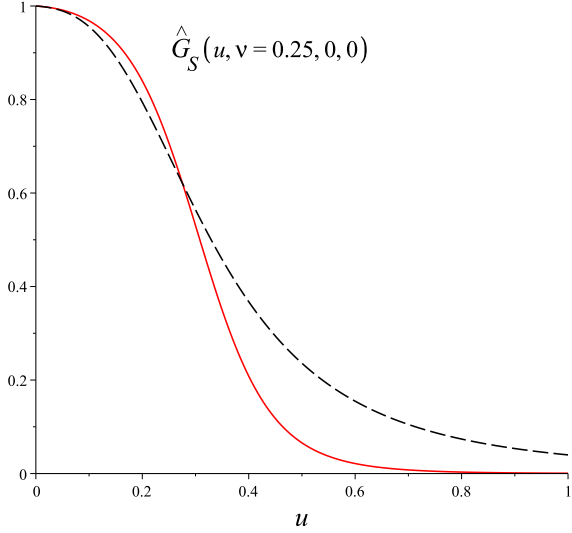} 
\end{center}
\caption{\label{fig:6}
The quantity $\hat G_S (u,\nu ,0,0)$ used in Damour-Nagar (dashed curve) \cite{Damour:2014sva} as well as the present determination (solid curve) are compared in the case $\nu=0.25$.
}
\end{figure}

One finds that after an initial small decrease from $c_3(0)=28.7012$ to $c_3(0.0041)=28.6175$, $c_3(u)$ then monotonically increases with $u$. It reaches the numerically  calibrated value of Refs. \cite{Damour:2014sva,Nagar:2015xqa}, namely $c_3^{\rm calibrated} = 44.786477$,  at $u=0.1593$ and then continues increasing towards large values (e.g. $c_3(0.5)=441.3976$). 
The inverse-resummed function $\hat G_S (u,\nu ,0,0)$ defined by inserting our result in  the second equation \eqref{GSnu} (with $\varphi_2(u) = 1$ and $O(\nu^2) \to 0$)  is shown in Fig. \ref{fig:6} for $\nu=0.25$ and is compared to the calibrated result of \cite{Damour:2014sva,Nagar:2015xqa}. Note that our results predict a faster fall-off of $\hat G_S$ in the strong-field domain. It will be interesting to explore the EOB application of this finding.

\smallskip

\appendix

\section{The Kerr case: an overview}

In the test-mass limit ($S_1=0=m_1$), i.e. in the Kerr case (with mass $m_2$ and spin $S_2$), 
the effective Hamiltonian reads
\beq
\label{H_eff_Kerr}
\hat H_{\rm eff}^{(0)}=\sqrt{A_K (1+p_\phi^2 u_{cK}^2)}+ 2u_K u_{cK}^2 p_\phi\hat a_2\,,
\eeq
where $u_K\equiv m_2/r$, $\hat a_2=a_2/m_2=S_2/m_2^2$ and
\beq
A_K(u_K,\hat a_2)=\frac{1-4u_{cK}^2}{1+2u_K}\,,\qquad  G_S^K(u_K,\hat a_2)=2u_Ku_{cK}^2\,,
\eeq
with
\begin{eqnarray}
r_{cK}(r,m_2,\hat a_2)^2&=& r^2+a_2^2 +\frac{2m_2a_2^2}{r}\nonumber\\
&=& \frac{m_2^2}{u_K^2}\left(1+\hat a_2^2 u_K^2 +2 \hat a_2 u_K^3 \right)\nonumber\\
u_{cK}(u_K,\hat a_2)&=& \frac{u_K}{\sqrt{1+\hat a_2^2 u_K^2(1+2u_K)}} \,.
\end{eqnarray}
Note also the expression of $A_K$ in terms of the Boyer-Lindquist coordinate $u_K = m_2/r$: 
\begin{eqnarray}
A_K (u_K) &= &1-2u_K + \frac{4 \hat a_2^2 u_K^4}{1+\hat a_2^2 u_K^2 (1+2u_K)} \nonumber  \\
&= &1-2u_K + 4\hat a_2^2 u_K^2 u_{cK}^2
\end{eqnarray}
The redshift $z_1 = \partial H_{\rm eff} / \partial \mu$, as a function of $u_K$ (hereafter we simply denote $u_K$ by $u$ and $u_{cK}$ by $u_c$) and $p_{\phi}$, reads 
\beq
\label{eq_z1_kerr}
z_1(u,p_{\phi},\hat a_2)=\frac{A_K(u,\hat a_2)}{\sqrt{A_K(u,\hat a_2)(1+p_\phi^2 u_c(u,\hat a_2)^2 )}}=\sqrt{\frac{A_K}{Q_K}}   \,,
\eeq
where
\beq
Q_K=1+p_\phi^2 u_c(u,\hat a_2)^2 \,.
\eeq
The circular value of the (dimensionless) angular momentum is 
\beq
\label{p_phi_kerr}
p_{\phi}^K(u,\hat a_2)= \frac{1-2\hat a_2 u^{3/2} +\hat a_2^2 u^2}{\sqrt{u} \sqrt{1-3u+2\hat a_2 u^{3/2}}}\,.
\eeq
The corresponding energy per unit mass, $\hat E = \hat H_{\rm eff}$, reads 
\beq
\hat E(u,\hat a_2)= \frac{1-2u+\hat a_2 u^{3/2}}{\sqrt{1-3u+2\hat a_2 u^{3/2}}}\,.
\eeq 
Substituting $p_{\phi} = p_{\phi}^K(u,\hat a_2)$  in the above  expression for $z_1^K$ yields the following \lq\lq on shell" relation
\beq
z_1^K(u,\hat a_2)= \frac{\sqrt{1-3u+2\hat a_2 u^{3/2}}}{1+\hat a_2 u^{3/2}}\,.
\eeq
The circular expression for the angular frequency parameter $m_2 \Omega^K = \partial \hat H_{\rm eff} / \partial p_{\phi}$ (as a function of $u$ after using $p_{\phi}=p_{\phi}^K(u,\hat a_2)$ given in \eqref{p_phi_kerr}) is   (Kepler's law, for a Kerr black hole)
\begin{eqnarray}
m_2\Omega^K(u,\hat a_2)&=& \frac{u^{3/2}}{1+\hat a_2 u^{3/2}}\nonumber\\
&=& \left( \hat a_2+u^{-3/2}\right)^{-1}\,.
\end{eqnarray}
It is worth to note the following relations
\begin{eqnarray}
m_2\Omega^K &=& \hat a_2 G_S^K+\frac{ A_K p_\phi^K u_{c}^2}{\sqrt{ A_K (1+p_\phi^K{}^2 u_{c}^2)}}\\
&=& \hat a_2 G_S^K+\sqrt{\frac{ A_K}{Q_K}} p_\phi^K u_{c}^2\\
&=&\hat a_2 G_S^K+ z_1^Kp_\phi^K u_{c}^2=u_{c}^2 \left( 2u \hat a_2 + z_1^Kp_\phi^K  \right)\,.\nonumber
\end{eqnarray}
Defining  
\beq
\Omega' \equiv \frac{\Omega}{1-a_2 \Omega}\,,\qquad m_2 \Omega ' \equiv \frac{m_2\Omega}{1-\hat a_2 m_2 \Omega}\,,
\eeq
one finds that this modified frequency satisfies the usual Kepler law
\beq
m_2 \Omega' =u^{3/2}\,.
\eeq
In other words, the modified dimensionless frequency parameter  
\beq
y'\equiv (m_2 \Omega')^{2/3}\,,
\eeq
is such that
\beq
u^{\rm circ} = y' \, .
\eeq
The explicit transformation $y \to y'$ reads
\begin{eqnarray}
\label{ypvsy}
y'&=&\frac{y}{(1-\hat a_2 y^{3/2})^{2/3}}\nonumber\\
&=&y\left[1+\frac23{\hat a}_2y^{3/2}+\frac59{\hat a}_2^2y^3+O(y^{9/2})\right]\,.
\end{eqnarray}
The inverse of this transformation is obtained by exchanging $y'$ and $y$ and ${\hat a}_2 \to - {\hat a}_2$, namely
\beq
y=\frac{y'}{(1+\hat a_2 y'{}^{3/2})^{2/3}}\,.
\eeq
Expressing $z_1(y')$ in terms of $y$ leads to
\begin{eqnarray}
z_1^K(y) &=& \frac{(1-3y'+2\hat a_2 y'^{3/2} )^{1/2}}{1+\hat a_2 y'^{3/2}}
\qquad {\rm (exact)}\nonumber\\
&=& \sqrt{1-3y}+\hat a_2 \left(2 y^{5/2}+3 y^{7/2}+\frac{27}{4} y^{9/2}\right. \nonumber\\
&& \left. +\frac{135}{8}y^{11/2} +\frac{2835}{64}y^{13/2}+\frac{15309}{128}y^{15/2} \right)\nonumber\\
&& +O(y^8,\hat a_2^2)\,.
\end{eqnarray}
Note that the equation
\beq
1-3y'+2\hat a_2 y'^{3/2} =0
\eeq
defines the light-ring for co-rotating circular geodesics. Table \ref{tab:7} lists light-ring values of $y'$ for representative values of $\hat a_2$.
\begin{table}
\centering
\caption{Light-ring position $y'$ for fixed values of $\hat a_2$. }
\begin{ruledtabular}
\begin{tabular}{cc}
$\hat a_2$ & $y'$ \cr
\hline
-1.0&0.25 \cr
-0.9& 0.2557369509 \cr 
-0.7& 0.2684314963\cr 
-0.5& 0.2831185829\cr 
0& 0.3333333333\cr
0.5& 0.4260220478\cr
0.7& 0.4966885468\cr
0.9& 0.6419084184\cr
1.0&1.0\cr
\end{tabular}
\end{ruledtabular}
\label{tab:7}
\end{table}
 
 Finally,  the  explicit expression, in a Kerr background, of the $G_{S_*}$-type spin-orbit coupling
 (defined in an arbitrary equatorial metric by Eq. \eqref{GSs}) reads
\begin{eqnarray}
 G_{S_*}^{\rm Kerr}(u,\hat a)&= &z_1^K u_c^2 \left[1-\sqrt{A}(1-\hat a^2 u^3)  \right] \nonumber \\
&+&\frac{u_c^4[(1+\hat a^2 u^2)^2-4\hat a^2 u^3]}{u(1+\sqrt{Q_K})}\,.
\end{eqnarray}
Note that this quantity differs from the ratio $R \equiv m_2 \Omega_{\rm SO}^K /p_\phi^K$ between the
dimensionless spin-orbit precession angular velocity \cite{Iyer:1993qa}
\beq
m_2 \Omega_{\rm SO}^K=\frac{1-\sqrt{1-3u+2\hat a_2 u^{3/2}}}{\hat a_2+u^{-3/2}}
\eeq
and the dimensionless angular momentum $p_{\phi}^K$. Indeed, the structure of the effective Hamiltonian \eqref{Heff} shows that 
$\Omega_{\rm SO}^K = \lim_{S_1 \to 0} \partial H_{\rm eff} / \partial S_1$ is (in the test-mass limit $m_1 \to 0$ with $S_1/m_1$ fixed) the sum of two contributions:
a contribution $\propto  G_{S_*} p_{\phi}$ and a contribution coming from the $S_1$ derivative of the orbital effective Hamiltonian $\sqrt{A(\mu^2+L^2/r_c^2)}$ [the
latter being even in spins, and therefore notably containing relevant terms of the form $ \sim S_1 (S_2 + S_2^3 + \ldots)$].
In the Schwarzschild limit, $G_{S_*}^{\rm Kerr}$ reduces to
\beq
G_{S_*}^{\rm Schw}(u)= \frac{3 u^3}{1+\frac{1}{\sqrt{1-3u}} }\,.
\eeq

\subsection*{Acknowledgments}

We are grateful to Abhay Shah for many informative discussions and for communicating us the corrected version of Table III in Ref. \cite{Shah:2012gu}.
D.B. thanks the Italian INFN (Naples) for partial support and IHES for hospitality during the development of this project.
All  the authors are grateful to ICRANet for partial support.


\begin{thebibliography}{99}

\bibitem{Buonanno:1998gg} 
A.~Buonanno and T.~Damour,
``Effective one-body approach to general relativistic two-body dynamics,''
Phys.\ Rev.\ D {\bf 59}, 084006 (1999)
[gr-qc/9811091].

\bibitem{Buonanno:2000ef} 
A.~Buonanno and T.~Damour,
``Transition from inspiral to plunge in binary black hole coalescences,''
Phys.\ Rev.\ D {\bf 62}, 064015 (2000)
[gr-qc/0001013].

\bibitem{Damour:2000we} 
T.~Damour, P.~Jaranowski and G.~Schaefer,
``On the determination of the last stable orbit for circular general relativistic binaries at the third postNewtonian approximation,''
Phys.\ Rev.\ D {\bf 62}, 084011 (2000)
[gr-qc/0005034].

\bibitem{Damour:2001tu} 
T.~Damour,
``Coalescence of two spinning black holes: an effective one-body approach,''
Phys.\ Rev.\ D {\bf 64}, 124013 (2001)
[gr-qc/0103018].

\bibitem{Schafer:2009dq} 
  G.~Schaefer,
  ``Post-Newtonian methods: Analytic results on the binary problem,''
  Fundam.\ Theor.\ Phys.\  {\bf 162}, 167 (2011)
  [arXiv:0910.2857 [gr-qc]].

\bibitem{Blanchet:2013haa} 
  L.~Blanchet,
  ``Gravitational Radiation from Post-Newtonian Sources and Inspiralling Compact Binaries,''
  Living Rev.\ Rel.\  {\bf 17}, 2 (2014)
  [arXiv:1310.1528 [gr-qc]].

\bibitem{Barack:2009ux} 
  L.~Barack,
  ``Gravitational self force in extreme mass-ratio inspirals,''
  Class.\ Quant.\ Grav.\  {\bf 26}, 213001 (2009)
  [arXiv:0908.1664 [gr-qc]].

\bibitem{Poisson:2011nh} 
  E.~Poisson, A.~Pound and I.~Vega,
  ``The Motion of point particles in curved spacetime,''
  Living Rev.\ Rel.\  {\bf 14}, 7 (2011)
  [arXiv:1102.0529 [gr-qc]].

\bibitem{Detweiler:2008ft} 
S.~L.~Detweiler,
 ``A Consequence of the gravitational self-force for circular orbits of the Schwarzschild geometry,''
Phys.\ Rev.\ D {\bf 77}, 124026 (2008)
[arXiv:0804.3529 [gr-qc]].

\bibitem{Blanchet:2009sd} 
  L.~Blanchet, S.~L.~Detweiler, A.~Le Tiec and B.~F.~Whiting,
  ``Post-Newtonian and Numerical Calculations of the Gravitational Self-Force for Circular Orbits in the Schwarzschild Geometry,''
  Phys.\ Rev.\ D {\bf 81}, 064004 (2010)
  [arXiv:0910.0207 [gr-qc]].

\bibitem{Damour:2009sm} 
T.~Damour,
``Gravitational Self Force in a Schwarzschild Background and the Effective One Body Formalism,''
Phys.\ Rev.\ D {\bf 81}, 024017 (2010)
[arXiv:0910.5533 [gr-qc]].

\bibitem{Blanchet:2010zd} 
  L.~Blanchet, S.~L.~Detweiler, A.~Le Tiec and B.~F.~Whiting,
  ``High-Order Post-Newtonian Fit of the Gravitational Self-Force for Circular Orbits in the Schwarzschild Geometry,''
  Phys.\ Rev.\ D {\bf 81}, 084033 (2010)
  [arXiv:1002.0726 [gr-qc]].

\bibitem{Barack:2010ny} 
  L.~Barack, T.~Damour and N.~Sago,
  ``Precession effect of the gravitational self-force in a Schwarzschild spacetime and the effective one-body formalism,''
  Phys.\ Rev.\ D {\bf 82}, 084036 (2010)
  [arXiv:1008.0935 [gr-qc]].

\bibitem{Keidl:2010pm} 
  T.~S.~Keidl, A.~G.~Shah, J.~L.~Friedman, D.~H.~Kim and L.~R.~Price,
  ``Gravitational Self-force in a Radiation Gauge,''
  Phys.\ Rev.\ D {\bf 82}, no. 12, 124012 (2010)
  [Phys.\ Rev.\ D {\bf 90}, no. 10, 109902 (2014)]
  [arXiv:1004.2276 [gr-qc]].

\bibitem{LeTiec:2011dp} 
  A.~Le Tiec, E.~Barausse and A.~Buonanno,
  ``Gravitational Self-Force Correction to the Binding Energy of Compact Binary Systems,''
  Phys.\ Rev.\ Lett.\  {\bf 108}, 131103 (2012)
  [arXiv:1111.5609 [gr-qc]].

\bibitem{Barausse:2011dq} 
E.~Barausse, A.~Buonanno and A.~Le Tiec,
``The complete non-spinning effective-one-body metric at linear order in the mass ratio,''
Phys.\ Rev.\ D {\bf 85}, 064010 (2012)
[arXiv:1111.5610 [gr-qc]].

 \bibitem{Shah:2012gu} 
  A.~G.~Shah, J.~L.~Friedman and T.~S.~Keidl,
  ``EMRI corrections to the angular velocity and redshift factor of a mass in circular orbit about a Kerr black hole,''
  Phys.\ Rev.\ D {\bf 86}, 084059 (2012)
  [arXiv:1207.5595 [gr-qc]].

\bibitem{Akcay:2012ea} 
S.~Akcay, L.~Barack, T.~Damour and N.~Sago,
``Gravitational self-force and the effective-one-body formalism between the innermost stable circular orbit and the light ring,''
Phys.\ Rev.\ D {\bf 86}, 104041 (2012)
[arXiv:1209.0964 [gr-qc]].

  \bibitem{Bini:2013zaa} 
  D.~Bini and T.~Damour,
  ``Analytical determination of the two-body gravitational interaction potential at the fourth post-Newtonian approximation,''
  Phys.\ Rev.\ D {\bf 87}, no. 12, 121501 (2013)
  [arXiv:1305.4884 [gr-qc]].

\bibitem{Dolan:2013roa} 
  S.~R.~Dolan, N.~Warburton, A.~I.~Harte, A.~L.~Tiec, B.~Wardell and L.~Barack,
  ``Gravitational self-torque and spin precession in compact binaries,''
  Phys.\ Rev.\ D {\bf 89}, 064011 (2014)
  [arXiv:1312.0775 [gr-qc]].

\bibitem{Shah:2013uya} 
  A.~G.~Shah, J.~L.~Friedman and B.~F.~Whiting,
  ``Finding high-order analytic post-Newtonian parameters from a high-precision numerical self-force calculation,''
  arXiv:1312.1952 [gr-qc].

\bibitem{Bini:2013rfa} 
  D.~Bini and T.~Damour,
  ``High-order post-Newtonian contributions to the two-body gravitational interaction potential from analytical gravitational self-force calculations,''
  Phys.\ Rev.\ D {\bf 89}, no. 6, 064063 (2014)
  [arXiv:1312.2503 [gr-qc]].
  
  \bibitem{Bini:2014nfa} 
  D.~Bini and T.~Damour,
  ``Analytic determination of the eight-and-a-half post-Newtonian self-force contributions to the two-body gravitational interaction potential,''
  Phys.\ Rev.\ D {\bf 89}, no. 10, 104047 (2014)
  [arXiv:1403.2366 [gr-qc]].

\bibitem{Bini:2014ica} 
  D.~Bini and T.~Damour,
   ``Two-body gravitational spin-orbit interaction at linear order in the mass ratio,''
  Phys.\ Rev.\ D {\bf 90}, no. 2, 024039 (2014)
  [arXiv:1404.2747 [gr-qc]].

\bibitem{Dolan:2014pja} 
  S.~R.~Dolan, P.~Nolan, A.~C.~Ottewill, N.~Warburton and B.~Wardell,
   ``Tidal invariants for compact binaries on quasicircular orbits,''
  Phys.\ Rev.\ D {\bf 91}, no. 2, 023009 (2015)
  [arXiv:1406.4890 [gr-qc]].
  
  \bibitem{Bini:2014zxa} 
  D.~Bini and T.~Damour,
  ``Gravitational self-force corrections to two-body tidal interactions and the effective one-body formalism,''
  Phys.\ Rev.\ D {\bf 90}, no. 12, 124037 (2014)
  [arXiv:1409.6933 [gr-qc]].

\bibitem{Bini:2015mza} 
  D.~Bini and T.~Damour,
  ``Analytic determination of high-order post-Newtonian self-force contributions to gravitational spin precession,''
  Phys.\ Rev.\ D {\bf 91}, no. 6, 064064 (2015)
  [arXiv:1503.01272 [gr-qc]].

\bibitem{Bini:2015bla} 
  D.~Bini and T.~Damour,
   ``Detweiler's gauge-invariant redshift variable: analytic determination of the nine and nine-and-a-half post-Newtonian self-force contributions,''
Phys.\ Rev.\ D {\bf 91}, 064050 (2015) 
  arXiv:1502.02450 [gr-qc].

\bibitem{Kavanagh:2015lva} 
  C.~Kavanagh, A.~C.~Ottewill and B.~Wardell,
  ``Analytical high-order post-Newtonian expansions for extreme mass ratio binaries,''
  Phys.\ Rev.\ D {\bf 92}, no. 8, 084025 (2015)
  [arXiv:1503.02334 [gr-qc]].

 \bibitem{Buonanno:2009qa} 
  A.~Buonanno, Y.~Pan, H.~P.~Pfeiffer, M.~A.~Scheel, L.~T.~Buchman and L.~E.~Kidder,
  ``Effective-one-body waveforms calibrated to numerical relativity simulations: Coalescence of non-spinning, equal-mass black holes,''
  Phys.\ Rev.\ D {\bf 79}, 124028 (2009)
  [arXiv:0902.0790 [gr-qc]].

\bibitem{LeTiec:2011bk} 
  A.~Le Tiec, A.~H.~Mroue, L.~Barack, A.~Buonanno, H.~P.~Pfeiffer, N.~Sago and A.~Taracchini,
  ``Periastron Advance in Black Hole Binaries,''
  Phys.\ Rev.\ Lett.\  {\bf 107}, 141101 (2011)
  [arXiv:1106.3278 [gr-qc]].

\bibitem{Damour:2011fu} 
  T.~Damour, A.~Nagar, D.~Pollney and C.~Reisswig,
  ``Energy versus Angular Momentum in Black Hole Binaries,''
  Phys.\ Rev.\ Lett.\  {\bf 108}, 131101 (2012)
  [arXiv:1110.2938 [gr-qc]].
  
   \bibitem{Damour:2012ky} 
  T.~Damour, A.~Nagar and S.~Bernuzzi,
  ``Improved effective-one-body description of coalescing nonspinning black-hole binaries and its numerical-relativity completion,''
  Phys.\ Rev.\ D {\bf 87}, 084035 (2013)
  [arXiv:1212.4357 [gr-qc]].
  
\bibitem{Hinder:2013oqa} 
  I.~Hinder, A.~Buonanno, M.~Boyle, Z.~B.~Etienne, J.~Healy, N.~K.~Johnson-McDaniel, A.~Nagar and H.~Nakano {\it et al.},
   ``Error-analysis and comparison to analytical models of numerical waveforms produced by the NRAR Collaboration,''
  Class.\ Quant.\ Grav.\  {\bf 31}, 025012 (2014)
  [arXiv:1307.5307 [gr-qc]].
  
\bibitem{Pan:2013rra} 
  Y.~Pan, A.~Buonanno, A.~Taracchini, L.~E.~Kidder, A.~H.~Mroué, H.~P.~Pfeiffer, M.~A.~Scheel and B.~Szilágyi,
   ``Inspiral-merger-ringdown waveforms of spinning, precessing black-hole binaries in the effective-one-body formalism,''
  Phys.\ Rev.\ D {\bf 89}, 084006 (2014)
  [arXiv:1307.6232 [gr-qc]].

\bibitem{Taracchini:2013rva} 
  A.~Taracchini {\it et al.},
  ``Effective-one-body model for black-hole binaries with generic mass ratios and spins,''
  Phys.\ Rev.\ D {\bf 89},  061502 (2014)
  [arXiv:1311.2544 [gr-qc]].

\bibitem{Pan:2013tva} 
  Y.~Pan {\it et al.},
   ``Stability of nonspinning effective-one-body model in approximating two-body dynamics and gravitational-wave emission,''
  Phys.\ Rev.\ D {\bf 89},  061501 (2014)
  [arXiv:1311.2565 [gr-qc]].

  \bibitem{Damour:2014afa} 
  T.~Damour, F.~Guercilena, I.~Hinder, S.~Hopper, A.~Nagar and L.~Rezzolla,
  ``Strong-Field Scattering of Two Black Holes: Numerics Versus Analytics,''
  arXiv:1402.7307 [gr-qc].
  
  \bibitem{Damour:2014sva} 
  T.~Damour and A.~Nagar,
   ``New effective-one-body description of coalescing nonprecessing spinning black-hole binaries,''
  Phys.\ Rev.\ D {\bf 90}, no. 4, 044018 (2014)
  [arXiv:1406.6913 [gr-qc]].

\bibitem{priv_com}
A.~G.~Shah, private communication.

\bibitem{shah_capra2015}
A.~G.~Shah, Capra conference 2015.

\bibitem{shah_MG14}
A.~G.~Shah, talk delivered at the MG14.

\bibitem{vandeMeent:2015lxa} 
  M.~van de Meent and A.~G.~Shah,
  ``Metric perturbations produced by eccentric equatorial orbits around a Kerr black hole,''
  Phys.\ Rev.\ D {\bf 92},  064025 (2015)
  [arXiv:1506.04755 [gr-qc]].

\bibitem{LeTiec:2011ab} 
  A.~Le Tiec, L.~Blanchet and B.~F.~Whiting,
  ``The First Law of Binary Black Hole Mechanics in General Relativity and Post-Newtonian Theory,''
  Phys.\ Rev.\ D {\bf 85}, 064039 (2012)
  [arXiv:1111.5378 [gr-qc]].

\bibitem{Blanchet:2012at} 
  L.~Blanchet, A.~Buonanno and A.~Le Tiec,
  ``First law of mechanics for black hole binaries with spins,''
  Phys.\ Rev.\ D {\bf 87}, no. 2, 024030 (2013)
  [arXiv:1211.1060 [gr-qc]].

\bibitem{Tiec:2015cxa} 
  A.~L.~Tiec,
  ``First Law of Mechanics for Compact Binaries on Eccentric Orbits,''
  arXiv:1506.05648 [gr-qc].
  
\bibitem{Damour:2007nc} 
  T.~Damour, P.~Jaranowski and G.~Schaefer,
  ``Hamiltonian of two spinning compact bodies with next-to-leading order gravitational spin-orbit coupling,''
  Phys.\ Rev.\ D {\bf 77}, 064032 (2008)
  [arXiv:0711.1048 [gr-qc]].

\bibitem{Balmelli:2015zsa} 
  S.~Balmelli and T.~Damour,
   ``A new effective-one-body Hamiltonian with next-to-leading order spin-spin coupling,''
  arXiv:1509.08135 [gr-qc].

\bibitem{Barausse:2009xi} 
  E.~Barausse and A.~Buonanno,
  ``An Improved effective-one-body Hamiltonian for spinning black-hole binaries,''
  Phys.\ Rev.\ D {\bf 81}, 084024 (2010)
  [arXiv:0912.3517 [gr-qc]].

\bibitem{Mano:1996mf} 
  S.~Mano, H.~Suzuki and E.~Takasugi,
  ``Analytic solutions of the Regge-Wheeler equation and the postMinkowskian expansion,''
  Prog.\ Theor.\ Phys.\  {\bf 96}, 549 (1996)
  [gr-qc/9605057].

\bibitem{Mano:1996vt} 
  S.~Mano, H.~Suzuki and E.~Takasugi,
  ``Analytic solutions of the Teukolsky equation and their low frequency expansions,''
  Prog.\ Theor.\ Phys.\  {\bf 95}, 1079 (1996)
  [gr-qc/9603020].

  \bibitem{Blanchet:2013txa}
  L.~Blanchet, G.~Faye and B.~F.~Whiting,
 ``Half-integral conservative post-Newtonian approximations in the redshift factor of black hole binaries,''
  Phys.\ Rev.\ D {\bf 89}, no. 6, 064026 (2014)
  [arXiv:1312.2975 [gr-qc]].

\bibitem{Damour:2015isa} 
  T.~Damour, P.~Jaranowski and G.~Sch\"afer,
  ``Fourth post-Newtonian effective one-body dynamics,''
  Phys.\ Rev.\ D {\bf 91}, no. 8, 084024 (2015)
  [arXiv:1502.07245 [gr-qc]].

\bibitem{Blanchet:1987wq} 
  L.~Blanchet and T.~Damour,
  ``Tail Transported Temporal Correlations in the Dynamics of a Gravitating System,''
  Phys.\ Rev.\ D {\bf 37}, 1410 (1988).

\bibitem{Johnson-McDaniel:2015vva} 
  N.~K.~Johnson-McDaniel, A.~G.~Shah and B.~F.~Whiting,
  ``Experimental mathematics meets gravitational self-force,''
  Phys.\ Rev.\ D {\bf 92}, no. 4, 044007 (2015)
  [arXiv:1503.02638 [gr-qc]].

\bibitem{Damour:2008qf} 
  T.~Damour, P.~Jaranowski and G.~Schaefer,
  ``Effective one body approach to the dynamics of two spinning black holes with next-to-leading order spin-orbit coupling,''
  Phys.\ Rev.\ D {\bf 78}, 024009 (2008)
  [arXiv:0803.0915 [gr-qc]].
  
\bibitem{Nagar:2011fx} 
  A.~Nagar,
  ``Effective one body Hamiltonian of two spinning black-holes with next-to-next-to-leading order spin-orbit coupling,''
  Phys.\ Rev.\ D {\bf 84}, 084028 (2011)
  [Phys.\ Rev.\ D {\bf 88}, no. 8, 089901 (2013)]
  [arXiv:1106.4349 [gr-qc]].

\bibitem{Barausse:2011ys} 
  E.~Barausse and A.~Buonanno,
  ``Extending the effective-one-body Hamiltonian of black-hole binaries to include next-to-next-to-leading spin-orbit couplings,''
  Phys.\ Rev.\ D {\bf 84}, 104027 (2011)
  [arXiv:1107.2904 [gr-qc]].

 \bibitem{Balmelli:2013zna} 
  S.~Balmelli and P.~Jetzer,
   ``Effective-one-body Hamiltonian with next-to-leading order spin-spin coupling for two nonprecessing black holes with aligned spins,''
  Phys.\ Rev.\ D {\bf 87}, no. 12, 124036 (2013)
  [Erratum-ibid.\ D {\bf 90}, no. 8, 089905 (2014)]
  [arXiv:1305.5674 [gr-qc]].

  \bibitem{Nagar:2015xqa} 
  A.~Nagar, T.~Damour, C.~Reisswig and D.~Pollney,
  ``Energetics and phasing of nonprecessing spinning coalescing black hole binaries,''
  arXiv:1506.08457 [gr-qc].

\bibitem{Iyer:1993qa} 
  B.~R.~Iyer and C.~V.~Vishveshwara,
 ``The Frenet-Serret description of gyroscopic precession,''
  Phys.\ Rev.\ D {\bf 48}, 5706 (1993)
  [gr-qc/9310019].

 \end{thebibliography}
\end{document}